\documentclass[10pt,journal,compsoc]{IEEEtran}

\ifCLASSOPTIONcompsoc
  \usepackage[nocompress]{cite}
\else
  \usepackage{cite}
\fi

\ifCLASSINFOpdf
\else
\fi

\usepackage[normalem]{ulem}
\usepackage{amsmath}
\usepackage{amssymb}
\usepackage{bbm}
\usepackage{xcolor}

\usepackage[numbers]{natbib}
\usepackage{graphicx}
\usepackage{subfigure}
\usepackage[skip=-1pt]{caption}
\usepackage{amsmath}
\usepackage{makecell}
\usepackage{booktabs}
\usepackage{amsfonts}
\usepackage{hyperref}
\usepackage{enumitem}
\usepackage{booktabs}
\usepackage{multirow}
\usepackage{mathtools}
\usepackage{multicol}
\usepackage{amsmath, bm}
\usepackage{stmaryrd}
\usepackage{longtable}

\DeclareMathOperator*{\argmax}{argmax}
\usepackage[ruled,vlined,english,onelanguage]{algorithm2e}
\usepackage{booktabs}
\usepackage{balance}

\usepackage{titlesec}
\titleformat{\paragraph}[runin]{\normalfont\normalsize\bfseries}{}{0pt}{}[\quad]
\titlespacing*{\paragraph}{0pt}{3.25ex plus 1ex minus .2ex}{1em}

\newcommand{\tabincell}[2]{\begin{tabular}{@{}#1@{}}#2\end{tabular}}
\newcommand{\new}[1]{\textcolor{black}{#1}}

\begin{document}

\title{Result Diversification in Search and Recommendation: A Survey}

\author{Haolun Wu\IEEEauthorrefmark{3}\thanks{\IEEEauthorrefmark{3}Equal contribution.}, 
        Yansen Zhang\IEEEauthorrefmark{3}, 
        Chen Ma\IEEEauthorrefmark{1}\thanks{\IEEEauthorrefmark{1}Corresponding author.},
        Fuyuan Lyu,
        Bowei He,
        Bhaskar Mitra,
        and Xue Liu~\IEEEmembership{Fellow,~IEEE}

\IEEEcompsocitemizethanks{
\IEEEcompsocthanksitem Haolun Wu, Fuyuan Lyu, and Xue Liu are with the School of Computer Science, McGill University, Montreal, Canada. E-mail: haolun.wu@mail.mcgill.ca; fuyuan.lyu@mail.mcgill.ca; xueliu@cs.mcgill.ca

\IEEEcompsocthanksitem Yansen Zhang, Chen Ma, and Bowei He are with the Department of Computer Science,
City University of Hong Kong, Hong Kong SAR. E-mail: yanszhang7-c@my.cityu.edu.hk; chenma@cityu.edu.hk; boweihe2-c@my.cityu.edu.hk

\IEEEcompsocthanksitem Bhaskar Mitra is with Microsoft Research, Montreal, Canada. E-mail: bhaskar.mitra@microsoft.com
}
}


\IEEEtitleabstractindextext{
\begin{abstract}
Diversifying return results is an important research topic in retrieval systems in order to satisfy both the various interests of customers and the equal market exposure of providers. There has been growing attention on diversity-aware research during recent years, accompanied by a proliferation of literature on methods to promote diversity in search and recommendation. However, diversity-aware studies in retrieval systems lack a systematic organization and are rather fragmented. In this survey, we are the first to propose a unified taxonomy for classifying the metrics and approaches of diversification in both search and recommendation, which are two of the most extensively researched fields of retrieval systems. We begin the survey with a brief discussion of why diversity is important in retrieval systems, followed by a summary of the various diversity concerns in search and recommendation, highlighting their relationship and differences. For the survey's main body, we present a unified taxonomy of diversification metrics and approaches in retrieval systems, from both the search and recommendation perspectives. In the later part of the survey, we discuss the open research questions of diversity-aware research in search and recommendation in an effort to inspire future innovations and encourage the implementation of diversity in real-world systems. We maintain an implementation for classical diversification metrics and methods summarized in this survey at \href{https://github.com/Forrest-Stone/Diversity}{https://github.com/Forrest-Stone/Diversity}.
\end{abstract}
}

\maketitle

\IEEEdisplaynontitleabstractindextext

\IEEEpeerreviewmaketitle

\section{Introduction}
\IEEEPARstart{W}{ith} the ever-growing volume of online information, users can easily access an increasingly vast number of online products and services.
To alleviate information overload and expedite the acquisition of information, information retrieval systems have emerged and begun to play important roles in modern society.
Search and recommendation are two of the most important applications of retrieval systems; both can be viewed as ranking systems that output an ordered list.
Search systems aim to retrieve relevant entities with respect to the information need(s) behind a query launched by users from a collection of resources.
Recommendation systems utilize the user-item interaction history to predict personalized user interests, hence recommending potentially satisfactory items to users.

For a long time, \textit{relevance} dominates the research in both search and recommendation, where the key is to measure if the system is able to retrieve those items that are regarded as ``relevant'' given part of the ground truth labels~\cite{DBLP:conf/uai/RendleFGS09, DBLP:conf/www/SarwarKKR01, he2020_lgcn}.
Although these systems are able to retrieve or recommend the most relevant items, they may jeopardize the utilities of stakeholders in the system.
Take the recommendation scenario as an example, systems with only high relevance have potential harms for both sides of the two most critical stakeholders: customers (the user side) and providers (the item side).
Customers generally suffer from the \textit{redundancy issue} that recommends redundant or very similar items, which may cause ``filter bubble''~\cite{filterbubble} and harm the customers' satisfaction in the long term.
For instance, a movie recommendation system keeping recommending Marvel action movies to a customer who once clicked ``\textit{The Batman}'' may block out the opportunity for her from observing other genres of movies. 
This does not necessarily indicate the customer only likes Marvel action movies.
It is highly possible that the customer has various interests but the recommender never explores other choices, thus jeopardizing the customer's long-term user experience.
Providers generally suffer from the \textit{exposure unfairness} due to the ``super-star''~\cite{superstar} economy phenomenon where a very small number of most popular items and providers take up an extremely large proportion of exposure to customers. 
Such unfairness may make those new-coming or less popular providers feel disappointed in attracting customers and finally quit the platforms.
Once only several most popular providers remain, monopoly is highly possible, harming a healthy marketplace and society.

From the perspective of search which is less concerned with personalization compared to recommendation, similar limitations occur when merely focusing on relevance.
Assuming there is an image search system, it will always retrieve images for ``jaguar vehicles'' when the query ``jaguar'' is entered.
Although it cannot be disputed that the output of this search system is significantly relevant to the input query, it cannot be considered ideal because the query ``jaguar'' has additional meanings, such as ``jaguar as an animal'', which are never retrieved.
Customers can be dissatisfied with the system for the inability to obtain the various desired information.
On the other hand, the system may also suffer from the similar \textit{exposure unfairness} described in the recommendation scenario for providers who offer jaguar animal pictures.
This is another example in the field of search that demonstrates how a singular focus on relevance can have negative effects on numerous stakeholders in a system.

Thus, in recent years, many criteria other than \textit{relevance} have gained tremendous attention in information retrieval systems, and \textit{diversity} is one of the paramount.
It has been recognized that diversified search and recommendation can not only increase the likelihood of satisfying various needs of customers in both the short term and long term, but also assist to increase item exposure, especially for those less popular providers~\cite{abs-1905-06589_Survey_DivRS, DBLP:conf/kdd/SunGZZRHGTYHC20}. 
Considering the critical role of diversity in maintaining a satisfactory and healthy information retrieval marketplace, we hereby offer a comprehensive review of definitions, metrics, and techniques of diversity studied in search and recommendation.

\textbf{Necessity of this Survey}. 
Although many papers have been published on this topic recently, to the best of our knowledge, none of them has provided an unified picture of diversity in both search and recommendation, as well as the corresponding diversity metrics and techniques.
We find that the usage of the terminology ``diversity'' in recent works is usually inconsistent across papers, without a clear claim as to which diversity perspective is emphasized.
In addition, some studies lacked an explanation of why they chose particular diversity criteria for measurement.
Given the growing awareness of the importance of diversity and the rapid development of diversity techniques in both search and recommendation, we believe our survey can provide a comprehensive summary and organization of the diversity concerns in these fields and offer future researchers a better comprehension of the current state-of-the-art and openness problems on this topic.

\textbf{Difference with Existing Surveys}.
A number of surveys in search and recommendation have been published recently, focusing on different perspectives. 
For instance, in the field of search,~\citet{AzadD19_Survey_QE} review the Query Expansion (QE) techniques and~\citet{Azzopardi21_Survey_Cog} summarizes the usage of cognitive bias.
In the field of recommendation,~\citet{Huang_2019_Survey_Privacy} provide a summary on privacy protection in recommendation systems and~\citet{abs-2010-03240_Survey_Bias} focus on bias and debias techniques. 
Some other well-cited surveys focus on more general problems in search and recommendation, such as~\cite{10.5555/222929_Survey_IR} and~\cite{ZhangYST19_Survey_DLRS}.
However, the perspective of diversity has
not been well reviewed in existing search and recommendation surveys. 
To the best of our knowledge, there exist several surveys on the diversity in recommendation~\cite{2017survey:DBLP:journals/kbs/KunaverP17, abs-1905-06589_Survey_DivRS, Diversity_survey}, but they do not systematically organize the diversity concerns in both search and recommendation, and the contents are not comprehensive or up-to-date.
To offer a comprehensive review of this topic, we make the following contributions in this survey:
\begin{itemize}
    \item Collecting the latest works and summarizing the types, metrics, and techniques of diversity in both search and recommendation systematically under a unified organization.
    \item Conducting a detailed analysis and presenting a taxonomy of existing diversity techniques, as well as discussing their strengths and drawbacks.
    \item Recognizing open research problems and discussing future directions to inspire more research on diversity in search and recommendation.
\end{itemize}

\begin{figure*}[t]
    \centering
    \includegraphics[width=0.95\linewidth]{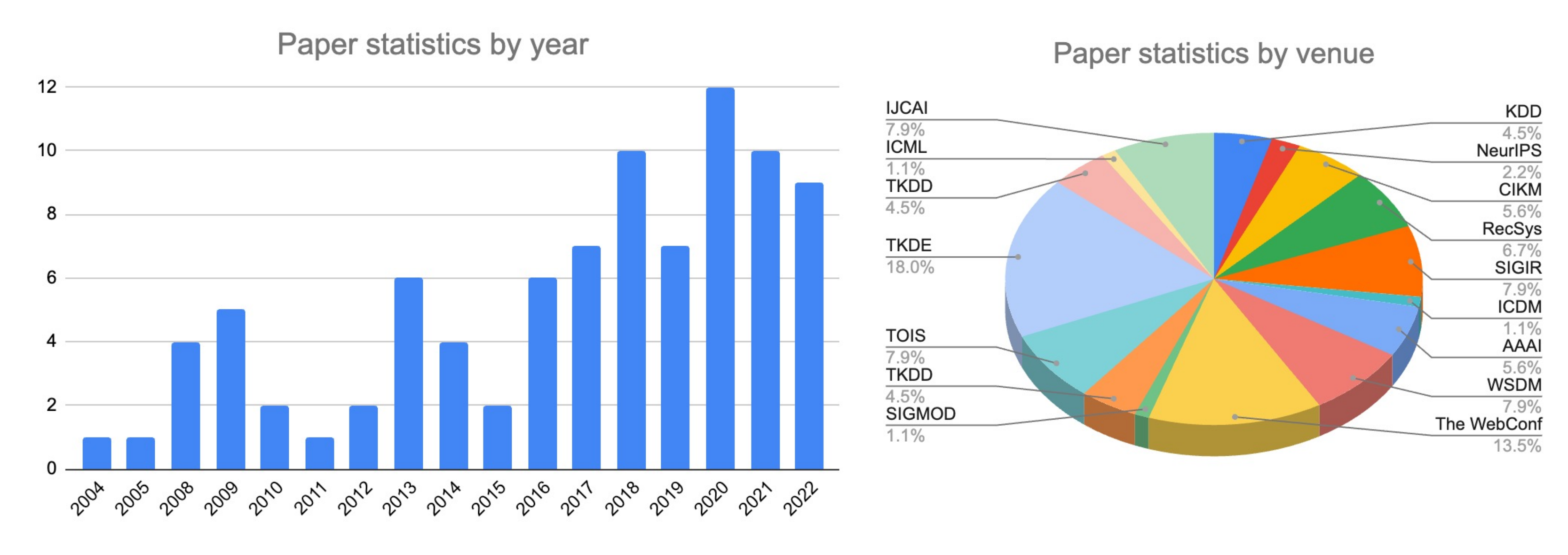}
    \caption{The statistics of publications related to diversity in search and recommendation with the publication year and venue.}
    \label{fig:statistics}
    \vspace{-4mm}
\end{figure*}

\textbf{Papers Collection}. 
We collect over 100 papers that analyze the diversity issues or propose novel techniques in search and recommendation.
We first search the related top-tier conferences and journals to find related work, including KDD, NeurIPS, CIKM, RecSys, ICDM, AAAI, WSDM, The WebConf, SIGIR, SIGMOD, TKDD, TKDE, TOIS, etc., with the keywords ``search'', ``recommendation'', ``ranking'' or ``retrieval'' combined with ``diversity'', ``serendipity'' or ``coverage'' till 2022. 
We then traverse the citation graph of the identified papers, retaining the papers that focus on diversity. 
Fig.~\ref{fig:statistics} illustrates the statistics of collected papers with the publication time and venue.

\textbf{Survey Audience and Organization}.
This survey is useful for researchers who are new to diversity problems and are looking for a guide to quickly enter this field, as well as those who wish to stay abreast of the most recent diversity strategies in search and recommendation.

The rest of the survey is organized as follows:
\begin{itemize}
    \item In Section~\ref{sec:div in search} and~\ref{sec:div in rec}, we summarize the categories and concerns of diversity in search and recommendation.
    \item In Section~\ref{sec:preliminaries}, we provide the background and preliminaries on search and recommendation systems, followed by listing the notations we used in this survey.
    \item In Section~\ref{sec:metrics}, we review the metrics of diversity generally used in search and recommendation, and systematically categorize them using a unified taxonomy.
    \item In Section~\ref{sec:offline-approach} and~\ref{sec:online-approach}, we review the approaches for enhancing diversity in search and recommendation, from both the offline and online perspectives.
    \item \new{In Section~\ref{sec:applicability}, we discuss the applicability of diversity metrics and approaches to various models.}
    \item In Section~\ref{sec:future}, we summarize the openness and future directions.
\end{itemize}
\section{Diversity in Search}
\label{sec:div in search}

Diversifying search results has received attention since the end of last century, where one of the earliest works is Maximal Marginal Relevance (MMR) proposed by~\citet{CarbonellG98MMR} in 1998.
Later,~\citet{ClarkeKCVABM08alphaNDCG} present a framework that systematically rewards novelty and diversity for measuring information retrieval systems, which promotes a series of works on diversity measurement and improvement in search.
As summarized by~\citet{RadlinskiBCJ09SIGIR_Forum} in the 2009 SIGIR Forum, diversity in search can be generally categorized into two classes based on whether the diversity is treated as uncertainty about the information need, or part of the information need.
These two concerns are named as (\romannumeral1) $\textit{extrinsic diversity}$ and (\romannumeral2) $\textit{intrinsic diversity}$ respectively, which are demonstrated as follows.

\subsection{Extrinsic Diversity}
Extrinsic diversity is related to the situation where uncertainty occurs in search, which can be further divided into the \textit{ambiguity} of the query meaning and the \textit{variability} of the user intent~\cite{RadlinskiBCJ09SIGIR_Forum}.
Generally, these two uncertainties co-occur in a search, as with the query ``jaguar''.
In other cases, even if there is no $\textit{ambiguity}$ in the query, the user intents may still contain $\textit{variability}$.
For instance, considering a query ``BioNTech, Pfizer vaccine'', a patient may seek more information on the vaccination's effect, whereas a doctor may be more concerned with the pharmaceutical ingredients and an entrepreneur may be more interested in the company operations of BioNTech.
The greater the ability of search results to encompass various query meanings and satisfy multiple user intents, the greater the extrinsic diversity.

\subsection{Intrinsic Diversity}
Different from extrinsic diversity which treats diversity as uncertainty about the information need, intrinsic diversity treats diversity as part of the information need itself, even given a single well-defined intent.
Under this definition, intrinsic diversity can be comprehended as the need for avoiding redundancy in the result lists, which is comparable to the novelty definition by~\citet{ClarkeKCVABM08alphaNDCG}.
The motivation for intrinsic diversity is intuitive: presuming the input query is ``jaguar as an animal'' with little ambiguity, users may anticipate the search results to contain images of different jaguars from diverse views and angles, rather than the same jaguar with the same view.
As such, the less redundancy in the search results, the greater the intrinsic diversity.

To clarify the distinction between extrinsic diversity and intrinsic diversity, the former is a response to a scenario with various search intents, whereas the latter is a response to a scenario with a single search purpose.
In real-world cases, both diversity concerns are significant in search and can be measured in a hierarchical and joint way. 
For instance, a search system may be expected to satisfy various information needs for diverse search intents, while avoiding redundancy for each specific one.

\section{Diversity in Recommendation}
\label{sec:div in rec}
\begin{figure*}[t]
    \centering
    \includegraphics[width=0.95\linewidth]{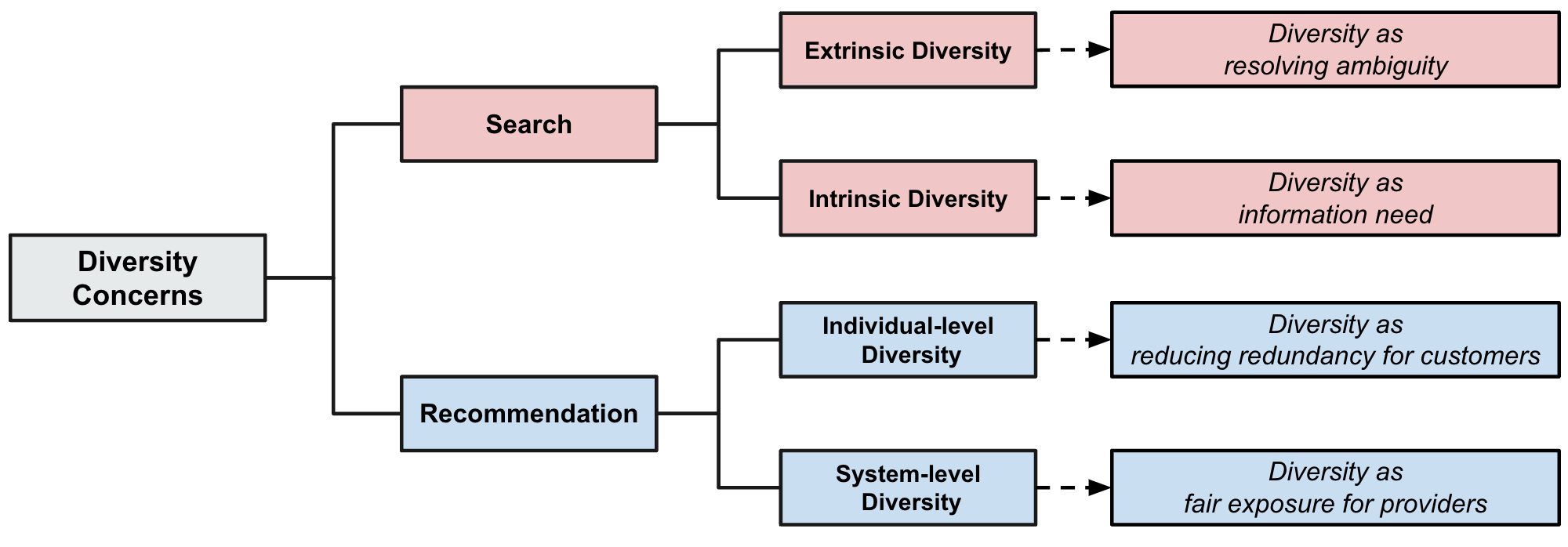}
    \caption{Diversity in search and recommendation. Pink boxes indicate that they were originally proposed and generally used in search, while blue boxes indicate that they are generally used in the recommendation.}
    \label{fig:div_concern}
    \vspace{-4mm}
\end{figure*}

As one of the most significant applications of information retrieval, the diversity in recommendation systems has also been explored.
Although the definition of diversity in search is also applicable in the field of recommendation, researchers study the diversity in this field from other perspectives.
There are generally two categories of diversity in recommendation across different works: (\romannumeral1) \textit{individual-level diversity} and (\romannumeral2) \textit{system-level (aggregated) diversity}.
Each diversity concern is relevant to one of the two most significant stakeholders: customers and providers, which represent the user side and the item side, respectively. 
The \textit{individual-level diversity} is relevant to the satisfaction of customers, while the \textit{system-level diversity} is relevant to the fairness of providers.
In this section, we offer a review and comparison of these two diversity concerns in recommendation systems.


%

\subsection{Individual-level Diversity}
The customer is one of the two most significant stakeholders in recommendation systems, whose satisfaction can be influenced by not only the recommendation relevance but also the diversity, serendipity, novelty, etc.
Therefore, one category of diversity in the recommendation, individual-level diversity, puts the customer at the core and is intended to quantify the degree to which recommended items are unique to each individual customer. 
As a result, the recommendation list for each individual is considered separately. 
From another perspective, individual-level diversity focuses on the problem of how to avoid recommending redundant (but still relevant) items to a customer given the previous recommendation list.
Thus, a higher degree of individual-level diversity can provide customers with the opportunity to view more novel items, thereby satisfying diverse demands and facilitating the exploration of various interests.

\subsection{System-level Diversity}
Rather than focusing on the redundancy of recommended items to each customer, system-level diversity reflects the ability of the entire system to recommend those less popular or hard-to-find items. 
Under this category, all customers are aggregated all together at a system level and the diversity measures the dissimilarity among all the recommended items the entire system had made.
A high degree of system-level diversity now indicates that the system can recommend a wide range of items rather than only those bestsellers or popular items, and is especially beneficial to those minority provider groups.
In other words, system-level diversity is relevant to exposure fairness among providers, which is important for maintaining a healthy marketplace.

\begin{figure}[t]
    \centering
    \includegraphics[width=1.0\linewidth]{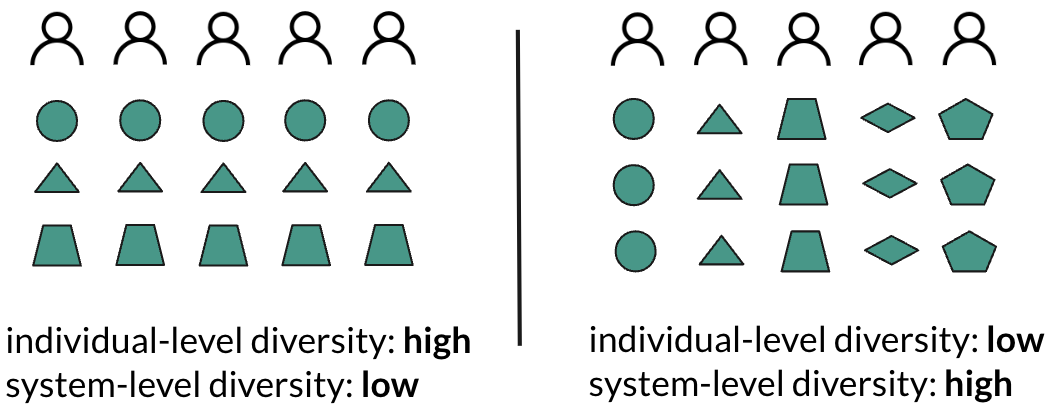}
    \caption{A toy example to show that the individual-level diversity and system-level diversity in the recommendation are different concerns with little overlap. In this illustration, different shapes refer to different categories. Consider a top-3 recommendation and assume that there is an extremely large number of users and items in the system (the same as the real-world scenarios). In case 1, the system recommends the same 3 categories of items to each user; in case 2, the system always recommends a unique category of items to each unique user. Therefore, in case 1, the individual-level diversity is high and the system-level diversity is low, while in case 2, the individual-level diversity is low and the system-level diversity is high.}
    \label{fig:toy_example}
    \vspace{-4mm}
\end{figure}

It is worth noting that individual-level diversity and system-level diversity address two distinct concerns with little overlap. 
System-level diversity is not a simple average of individual-level diversity across all customers.
It is conceivable for a system to have a high degree of individual-level diversity but a low degree of system-level diversity, and vice versa. 
For example, if the system recommends to all users the same five best-selling items that are not similar to each other, the recommendation list for each user is diverse (i.e., high individual-level diversity), but only five distinct items are recommended to all users and purchased by them (i.e., resulting in low system-level diversity or high sales concentration). 
In the other case, if the system recommends the same and unique category of items to each user, then the individual-level diversity is low, while the system-level diversity can be high. 
A toy example is provided in Fig.~\ref{fig:toy_example}.

\vspace{-2mm}
\section{Notations}
\label{sec:preliminaries}
The notations we used in this paper are shown in Table~\ref{table:notation1}. 
\vspace{-2mm}
\begin{table}[t]
\centering
\caption{\label{tab:formulation}\new{Description of Notations.}}
\vspace{1mm}
\resizebox{0.48\textwidth}{!}{
\begin{tabular}{ll}
\toprule
    \textbf{Notation} & \textbf{Description}  \\
\midrule
$\mathcal{U}, \mathcal{D}$ & The set of all users and items\\
$\mathcal{D}_u$ & The set of interacted items of user $u$\\
$u, d$ & An individual user / item\\
$\Theta$ & Learnable embeddings of users, items, and subtopics (if applicable)\\
$\mathbf{M}$ & The interaction matrix between users and items\\
$\sigma$ & A ranking list of items \\
$\sigma^{i:j}$ & The list of items from position $i$ to position $j$ extracted from $\sigma$ \\
$\sigma^{-1}(d)$ & The ranking position of item $d$ in $\sigma$ \\
$n_S$ & The total number of subtopics (categories)\\
$\mathcal{S}(d)$ & The set of subtopics covered by item $d$\\
$c_s^{l}$ & The number of items covering subtopic $s$ in list $l$\\
$e_i$ & The exposure of item $d_i$ in the entire system\\
$o(d|u)$ & The score of item $d$ with respect to user $u$ \\
$p(s|u)$ & The user $u$'s interest in subtopic $s$\\
$p(d|s)$ & The relatedness of items $d$ to subtopic $s$\\
\bottomrule
\end{tabular}}
\vspace{-4mm}
\label{table:notation1}
\end{table}

\section{Metrics of Diversity in Search and Recommendation}
\label{sec:metrics}

\begin{figure*}[t]
    \centering
    \includegraphics[width=0.95\linewidth]{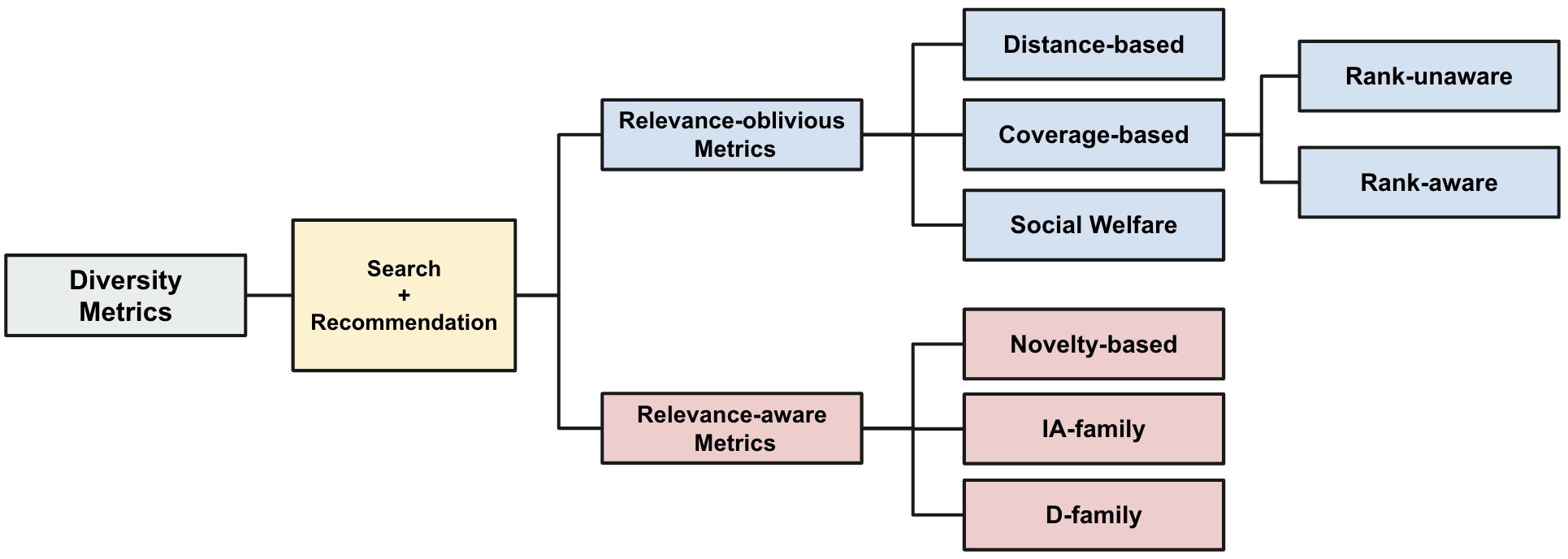}
    \caption{Diversity metrics in search and recommendation. We unify the metrics in one unified classification since all metrics can be theoretically used in both fields. Metrics in pink boxes indicate that they were originally proposed and generally used in search, while those in blue boxes indicate that they are generally used in recommendation.}
    \label{fig:div_metric}
    \vspace{-4mm}
\end{figure*}

Although many diversity metrics were proposed separately in either the field of search or recommendation, they can actually be applied interchangeably in both fields, since they all commonly aim to measure the dissimilarity and non-redundancy among a list of items.
\new{Coming up with a unified classification rationale for diversity metrics is necessary but not easy since diversity can be measured from multiple perspectives in search and recommendation. 
For example, some~\cite{DBLP:conf/recsys/ZhangH08, DBLP:conf/www/ZieglerMKL05} adopt the dissimilarity to define diversity and use the Intra-List Average Distance (ILAD) to measure diversity.
Others~\cite{ClarkeKCVABM08alphaNDCG, ChengWMSX17Acc_Diverse} consider the item position and relatedness to different subtopics to determine the diversity of the recommendation list and use $\alpha$-NDCG to measure diversity.
Some other researchers~\cite{AgrawalGHI09Intent, amigo2018axiomatic} define diversity in a way where the relevance also being inherently considered. 
For instance,~\citet{AgrawalGHI09Intent} state the problem of result diversification as: ``\textit{Suppose users only consider the top $k$ returned results of a search engine. Our objective is to maximize the probability that the average user finds at least one useful result within the top $k$ results}''. 
The word ``\textit{useful}'' here is related to relevance inherently. 
The diversity metrics~\citet{AgrawalGHI09Intent} later defined (i.e., the Intent-aware family (IA-family) of metrics) are also related to relevance themselves, since the metrics have captured user preference on recommended items based on the user intents.}

In this survey, \new{motivated by prior analysis and studies~\cite{amigo2018axiomatic, 2017survey:DBLP:journals/kbs/KunaverP17, Neil2011DiveRS}}, we summarize the metrics of diversity in both fields under one unified taxonomy as follows.
We first categorize diversity metrics into two classes based on whether the relevance of items~\footnote{To clarify, the term ``\textit{item}'' in the rest of this paper can refer to both ``\textit{entities retrieved from search systems}'' and ``\textit{goods displayed by recommendation systems}''.} to the user (query) is taken into consideration: (\romannumeral1) \textbf{Relevance-oblivious Diversity Metrics} and (\romannumeral2) \textbf{Relevance-aware Diversity Metrics}.
We further classify these metrics into sub-classes in a more fine-grained manner.
A summary table is maintained to highlight the applicability of all these metrics in either search, recommendation, or both.
These metrics and corresponding works are summarized in Table~\ref{tab:metrics}.

\begin{table*}[h]
\caption{A lookup table for the publications proposing or using different diversity metrics in search and recommendation.}
\label{tab:metrics}
\vspace{1mm}
\centering
\resizebox{\linewidth}{!}{
\begin{tabular}{l|l|l|l|l}
\Xhline{2\arrayrulewidth}
\multicolumn{4}{l|}{\textbf{Diversity Metrics}}                       &       \textbf{Related Work}                                                                                                                                                                                                                                                                           \\ \hline
\multirow{12}{*}{\makecell{Relevance-oblivious\\Diversity Metrics}} & \multirow{3}{*}{Distance-based} & \multicolumn{2}{c|}{Cosine Diversity Distance} & \tabincell{l}{\cite{DBLP:conf/recsys/ZhangH08,DBLP:conf/www/ZieglerMKL05,NIPS2018:DBLP:conf/nips/ChenZZ18,DBLP:conf/kdd/HuangWZX21,CIKM2020:DBLP:conf/cikm/ChenRC0R20,ChengWMSX17Acc_Diverse,li2020directional,GanLC20DPP4,Chen2017DPP1,liang2021enhancing,DBLP:conf/wsdm/ParaparR21} \\ \cite{WasilewskiH16Diverse_LTR,vargas2011intent,DBLP:conf/wsdm/StamenkovicKAXK22,chen2021multi,han2019geographic}} \\
\cline{3-5} 
& & \multicolumn{2}{c|}{Jaccard Diversity Distance} & \cite{DBLP:conf/sigir/GongZC0B0YQ22,DBLP:conf/recsys/TsukudaG19} \\
\cline{3-5} 
& & \multicolumn{2}{c|}{Gower Diversity Distance} & \cite{Haritsa09Gower} \\
\cline{2-5} 
 & \multirow{7}{*}{Coverage-based} & \multirow{4}{*}{Rank-unaware} & P-Coverage & \cite{DBLP:journals/tois/HerlockerKTR04,DBLP:conf/recsys/GeDJ10,DBLP:conf/recsys/PaudelHB17,DBLP:conf/wsdm/StamenkovicKAXK22,DBLP:conf/sigir/BalloccuBFM22, DBLP:journals/tkde/AdomaviciusK12,raman2014understanding} \\
 \cline{4-5}
 & & & C-Coverage & \cite{DBLP:journals/tois/HerlockerKTR04,DBLP:conf/recsys/GeDJ10,Wilhelm2018DPP2,13KwonHH20,ZhengGCJL21DGCN-pre,han2019geographic} \\
 \cline{4-5}
  & & & S-Coverage & \cite{DBLP:conf/recsys/GeDJ10,CIKM2020:DBLP:conf/cikm/ZhouAK20,he2019diversity,ChengWMSX17Acc_Diverse,liang2021enhancing} \\
  \cline{4-5}
  & & & E-Coverage & \cite{DBLP:journals/tkde/YinHLZ13, DBLP:journals/tkde/LiY13} \\
  \cline{3-5}
 & & \multirow{3}{*}{Rank-aware} & S-RR@100\% & \cite{Zhai_Subtopic} \\
 \cline{4-5}
 & & & S-Recall@K & \cite{CIKM2020:DBLP:conf/cikm/QinDW20,Zhai_Subtopic, DBLP:journals/tkde/LiangCRR16,han2019geographic,liang2017search} \\
 \cline{4-5}
  & & & S-Precision@K & \cite{Zhai_Subtopic} \\
 \cline{2-5}
 & \multirow{2}{*}{\hfil Social Welfare} & \multicolumn{2}{c|}{SD Index} & \cite{CIKM2020:DBLP:conf/cikm/ZhouAK20,simpson1949measurement} \\ 
\cline{3-5} 
& & \multicolumn{2}{c|}{Gini Index} & \cite{antikacioglu2017post,DBLP:conf/recsys/Sanz-CruzadoC18,ZhengGCJL21DGCN-pre,GiniIndex} \\
\hline
\multirow{5}{*}{\makecell{Relevance-aware\\Diversity Metrics}} & \multirow{3}{*}{Novelty-based} & \multicolumn{2}{c|}{$\alpha$-nDCG@K} & \tabincell{l}{\cite{ClarkeKCVABM08alphaNDCG,CIKM2020:DBLP:conf/cikm/QinDW20,CIKM2020:DBLP:conf/cikm/ZhouAK20,DBLP:conf/recsys/ParaparR21,ChengWMSX17Acc_Diverse,LiZZZL17DiverseMF,SantosMO10xquard,vargas2011intent,DBLP:conf/aaai/Yu22, DBLP:journals/tkde/LiangCRR16, DBLP:journals/tkde/JiangDZNYW18} \\
\cite{han2019geographic,liang2017search,cai2016diversifying}} \\
\cline{3-5} 
& & \multicolumn{2}{c|}{NRBP} & \cite{CIKM2020:DBLP:conf/cikm/QinDW20, DBLP:journals/tkde/JiangDZNYW18,liang2017search,cai2016diversifying} \\
\cline{3-5} 
& & \multicolumn{2}{c|}{nDCU@K} & \cite{YangLLHKR07nDCU, YangL09EGU} \\
\cline{2-5} 
 & IA-family & \multicolumn{2}{c|}{M-IA} & \cite{AgrawalGHI09Intent,CIKM2020:DBLP:conf/cikm/QinDW20,LiZZZL17DiverseMF,SantosMO10xquard,vargas2011intent,DBLP:conf/aaai/Yu22, DBLP:journals/tkde/LiangCRR16, DBLP:journals/tkde/JiangDZNYW18,sakai2020retrieval,liang2017search,cai2016diversifying} \\ 
\cline{2-5}
 & D-family & \multicolumn{2}{c|}{D-M, D\#-M} & \cite{sakai2012evaluation,sakai2011evaluating,sakai2020retrieval,DBLP:conf/ntcir/SakaiCSRDL10, DBLP:journals/tkde/LiangCRR16, DBLP:journals/tkde/WangWDSZ18} \\ 

\Xhline{2\arrayrulewidth}
\end{tabular}%
}
\vspace{-4mm}
\end{table*}

\subsection{Relevance-oblivious Diversity Metrics}
The relevance-oblivious metrics do not take the relevance between items and user (query) into consideration, while merely focus on the diversity measurement of a ranking list itself. 
We further categorize these metrics into the following three sub-classes: \textit{Distance-based Metrics}, \textit{Coverage-based Metrics}, and \textit{Social Welfare Metrics}.

\vspace{-2mm}
\subsubsection{Distance-based Metrics}
One of the most widely used metrics for measuring diversity is the distance-based metric. 
As the name says, it evaluates the diversity by calculating the pair-wise distance among all the items in a list, where a smaller distance value indicates a poorer diversity. 
Given a specific criterion for computing the pair-wise distance, most works follow the \textbf{ILAD} and \textbf{ILMD} (short for \textbf{I}ntra-\textbf{L}ist \textbf{A}verage \textbf{D}istance and \textbf{I}ntra-\textbf{L}ist \textbf{M}inimal \textbf{D}istance) paradigm for obtaining the diversity value of the item list.
These two paradigms originated from paper~\cite{DBLP:conf/recsys/ZhangH08} for measuring the diversity of a list, which extended the metrics of intra-list similarity proposed by~\citet{DBLP:conf/www/ZieglerMKL05}. 
We denote $\sigma$ as a retrieval or recommendation list, $\text{dis}_{ij}$ as the distance between item $d_i$ and $d_j$.
Then the ILAD can be defined as the average dissimilarity of all pairs of items in the list, while the ILMD can be defined analogously.
\begin{equation}        \text{ILAD}=\mathop{\text{mean}}\limits_{d_i,d_j\in \sigma,i\neq j}\text{dis}_{ij}, \quad \text{ILMD}=\mathop{\text{min}}\limits_{d_i,d_j\in \sigma,i\neq j}\text{dis}_{ij}.
\label{eq:ilad}
\end{equation}
In some specific applications such as sequential recommendation, items are displayed as a sequence to the user, and the diversity is only required for $w$ successive items~\cite{NIPS2018:DBLP:conf/nips/ChenZZ18}. 
We denote $\sigma^{-1}(d)$ as the rank position of item $d$ in $\sigma$, then we can obtain two variants of ILAD and ILMD as ILALD and ILMLD (short for \textbf{I}ntra-\textbf{L}ist \textbf{A}verage \textbf{L}ocal \textbf{D}istance and \textbf{I}ntra-\textbf{L}ist \textbf{M}inimal \textbf{L}ocal \textbf{D}istance), which can be defined as:
\begin{align}
\text{ILALD}&=\mathop{\text{mean}}\limits_{d_i,d_j\in \sigma,i\neq j, \atop |\sigma^{-1}(d_i)-\sigma^{-1}(d_j)|\leq w}\text{dis}_{ij},\\
\text{ILMLD}&=\mathop{\text{min}}\limits_{d_i,d_j\in \sigma,i\neq j, \atop |\sigma^{-1}(d_i)-\sigma^{-1}(d_j)|\leq w}\text{dis}_{ij}.
\end{align}

    
    

Based on different ways to calculate the pair-wise distance $\text{dis}_{ij}$, several specific metrics are summarized below.

\begin{itemize}

    \item \textbf{Cosine Diversity Distance}.
    The most traditional and widely adopted way for defining the pair-wise distance is to use the cosine similarity between item embeddings, where  $\text{dis}_{ij}$ is generally defined as $\text{dis}_{ij}=1-cos\langle\vec{d}_i, \vec{d}_j\rangle$, where $cos\langle\cdot, \cdot\rangle$ refers to the cosine similarity. 
    One of the primary advantages of cosine similarity is its simplicity, especially for sparse vectors — only non-zero entries need to be considered.
    This is also how the original ILAD and ILMD proposed by~\citet{DBLP:conf/www/ZieglerMKL05} define the pair-wise distance.
    After computing the cosine distance between any pair of items within the list, the Cosine diversity distance of the whole list can be obtained using the paradigm in Eq.~\ref{eq:ilad}.

    \item \textbf{Jaccard Diversity Distance}.
    Proposed by~\citet{YuLA09JaccardDD}, the Jaccard diversity distance is calculated similarly to a standard Jaccard index paradigm. 
    However, the exact distance is not computed based on item embeddings, but relies on $\textit{explanation}$ between user-item pairs.
    The $\textit{explanation}$ is defined differently given different recommendation models. 
    If an item $d_i$ is recommended to user $u$ by a content-based strategy, then an $\textit{explanation}$ for recommendation $d_i$ is defined as:
    \begin{equation}
        \text{Expl}(u, d_i)=\{d_j\in\mathcal{D}|\text{sim}(d_i, d_j)>0 \land d_j\in\mathcal{D}_u\}.
    \end{equation}
    Thereafter, the Jaccard diversity distance (JDD) between two items recommended to a specific user can be defined using the pre-computed $\textit{explanation}$:
    \begin{equation}
        \text{JDD}(d_i, d_j|u)=1-\frac{|\text{Expl}(u, d_i)\cap\text{Expl}(u, d_j)|}{|\text{Expl}(u, d_i)\cup\text{Expl}(u, d_j)|}.
    \end{equation}
    Then the Jaccard diversity distance of the whole list can be defined  as Eq.~\ref{eq:ilad}.

    \item \textbf{Gower Diversity Distance}. Another metric belonging to the distance-based category is the Gower diversity distance, proposed by~\citet{Haritsa09Gower}, focusing on retrieving $K$ nearest and diversified items with respect to a given query.

    Motivated by the Gower coefficient~\cite{Gower_origin}, they define the distinction between two items as a weighted average of the respective attribute value differences.
    Assuming $\delta_k$ is the difference of the $k^{th}$ attribute between two items and $w_k$ is the corresponding weight on that attribute, then the Gower diversity distance (GDD) between two items $d_i$ and $d_j$ can be defined as:
    \begin{equation}
        \text{GDD}(d_i, d_j)=\sum_{k}w_k\cdot \delta_k (d_i, d_j).
    \end{equation}
    The rest of the computation over a whole list follows the same paradigm in Eq.~\ref{eq:ilad}.
    
\end{itemize}

\subsubsection{Coverage-based Metrics}
Coverage-based metrics, popular for diversity measurement in search and recommendation, are often designed to quantify the breadth of \textit{subtopics}~\footnote{The term ``\textit{subtopic}'' originates from information retrieval, indicating the presence of multiple themes or keywords relevant to the input query. 
For consistency, we use ``\textit{subtopic}'' to represent (\romannumeral1) \textit{category of items}, (\romannumeral2) \textit{aspect of queries}, and (\romannumeral3) \textit{intent of users} in this survey.} within a list of unique items.
Depending on whether item ranks matter, we classify metrics as \textit{rank-unaware} or \textit{rank-based}. 
The first category disregards the ranking positions of items.

\vspace{-2mm}
\paragraph{\textbf{Rank-unaware}.}
Rank-unaware metrics are similar to the conventional metrics on accuracy in search and recommendation (e.g., Precision@K and Recall@K) since both of them will not be influenced by the rank positions of items in a given list. 
Depending on the coverage of ``what'' they measure, we can classify these metrics into three sub-classes: P-Coverage, C-Coverage, and S-Coverage.

\begin{itemize}
    \item \textbf{P-Coverage} (short for \textbf{P}rediction Coverage). The measure of prediction coverage is the number of unique items for which the predictions can be formulated as a proportion of the total number of items~\cite{DBLP:journals/tois/HerlockerKTR04,DBLP:conf/recsys/GeDJ10}. 
    We denote $\mathcal{D}$ as the set of all available items, $\mathcal{D}_p$ as the set of items for which a prediction can be provided. 
    Then the P-Coverage can be defined as follows:
    \begin{equation} 
        \text{P-Coverage}=\frac{\left|\mathcal{D}_p\right|}{\left|\mathcal{D}\right|}.
    \end{equation}
    The construction of $\mathcal{D}_p$ is highly dependent on the task formulation and chosen models.
    For instance, paper~\cite{DBLP:conf/recsys/GeDJ10} mentioned that some collaborative filtering systems are just able to make predictions for items that have more than a fixed number of ratings assigned. 
    In such a case, $\mathcal{D}_p$ can be considered as the set of items for which the number of available ratings meets the requirement. 
    P-Coverage generally focuses on the system level, and is hardly used for measurement on a single or several ranking lists.
    
    From another view, we can also understand the P-Coverage as the system's ability to address the ``cold-start'' problem.
    However, as more and more research on ``cold-start'' problem emerges, most models are capable of making predictions for those items even with very few interactions.
    Thus, P-Coverage is not widely used in search and recommendation.
    
    \item \textbf{C-Coverage} (short for \textbf{C}atalog Coverage). In order to quantify the proportion of unique items that can be retrieved or recommended in the system, the catalog coverage metric directly focuses on the output result list and generally considers the union of all lists produced during the measurement time~\cite{DBLP:journals/tois/HerlockerKTR04,DBLP:conf/recsys/GeDJ10}.
    C-Coverage can be used to measure the diversity of either a single ranking list or a group of lists, but is more widely used at a system level.
    Assuming there are $N$ lists, the metric can be formulated as follows, \new{where $\text{set}(\cdot)$ is to convert a ranking list to a set}:
    \begin{equation} 
        \text{C-Coverage}=\frac{\left|\bigcup_{i=1}^ N\text{set}(\sigma_i)\right|}{\left|\mathcal{D}\right|}.
    \end{equation}

    \item \textbf{S-Coverage} (short for \textbf{S}ubtopic-Coverage). This metric is one of the most widely used measurements for diversity in search and recommendation~\cite{DBLP:conf/recsys/GeDJ10, he2019diversity}. 
    Differing from C-Coverage, which focuses on the items themselves, S-Coverage considers the variety and richness of different item categories or genres within the list. 
    This aligns more naturally with human perceptions than distance-based metrics, as people typically do not compute pair-wise distances based on embeddings to assess list diversity, but rather identify whether diverse topics are as frequent as possible. 
    S-Coverage can be gauged on either a single list or multiple lists, corresponding respectively to individual-level and system-level diversity measurements. 
    If we denote $N$ as the number of lists in consideration, $\mathcal{S}(d)$ as the set of subtopics covered by item $d$, and $n_S=\left|\bigcup_{d\in\mathcal{D}}\mathcal{S}(d)\right|$ as the total number of subtopics, then S-Coverage can be expressed as follows:
    \begin{equation} 
    \text{S-Coverage}=\frac{\left|\bigcup_{i=1}^ N\left(\bigcup_{d\in \sigma_i}\mathcal{S}(d)\right)\right|}{n_S}.
    \end{equation}

    \item \textbf{Density} is another member in the family of coverage-based metrics, which derives from the network science and is widely applied on graphs and information networks~\cite{DBLP:journals/tkde/YinHLZ13, DBLP:journals/tkde/LiY13}.
    The density of a graph is defined as the number of edges (excluding self-links) presenting in the network divided by the maximal possible number of edges in the network. 
    We re-name it as \textbf{E-Coverage} (short for \textbf{E}dge-Coverage) in our survey. 
\end{itemize}

\vspace{-3mm}
\paragraph{\textbf{Rank-aware}.}
It has been realized in many works that users do not provide all items in a ranking list with the same amount of attention due to users' patience may decay exponentially as they browse deeper through a list.
As a result, those items ranked higher (i.e., at the top of the list) may receive more exposure.
Thus, when considering relevance, many metrics have been proposed to offer a higher score for ranking relevant items at the top of a list, such as the normalized discounted cumulative gain (nDCG).

A similar idea is also applicable when considering the diversity of a list: a user may feel the list is redundant if those items ranked in top positions are similar to each other, even if there are many diverse items in later positions.
This nuance cannot be captured by prior described metrics since they are invariant when the ranks of items change.
Here, we summarize the rank-based metrics on measuring coverage-based diversity, where most of them are defined upon the conventional metrics for measuring accuracy in search and recommendation.
These metrics care about not only how diverse the items are but also what locations they appear. 

\begin{itemize}
    \item \textbf{S-RR@100\%} (short for \textbf{S}ubtopic-\textbf{R}eciprocal \textbf{R}ank@100\%).
    This metric is proposed in~\cite{Zhai_Subtopic} for evaluating the diversity of solutions for subtopic retrieval problems. 
    The subtopic retrieval problem is originally concerned with finding documents that cover many different subtopics given a query of keywords.
    S-RR@100\% is a variation to Reciprocal Rank (RR), defined as the inverse of the rank position on which a complete coverage of subtopics is obtained.
    Thus, the output value of this metric cannot be smaller than the total number of different subtopics.
    Using the same notation as before, S-RR@100\% can be defined as:
    \begin{equation}
    \text{S-RR@100\%}=\min\limits_k\left(\left|\bigcup_{i=1}^k \mathcal{S}(d_i)\right|=n_S\right).
    \end{equation}
    
    \item \textbf{S-Recall@K} (short for \textbf{S}ubtopic-Recall@K).
    As the name says, this metric is a variation of the Recall@K metric that is widely used for measuring relevance in search and recommendation. 
    S-Recall@K is also proposed in~\cite{Zhai_Subtopic} and can be defined as the percentage of subtopics covered by the first $k$ items given a retrieved or recommendation list:
    \begin{equation}
    \text{S-Recall@K}=\frac{\left|\bigcup_{i=1}^k \mathcal{S}(d_i)\right|}{n_S}.
    \end{equation}

    \item \textbf{S-Precision@K} (short for \textbf{S}ubtopic-Precision@K).
    Analogous to S-Recall@K, this metric is a variation of the Precision@K. 
    It can be defined as below:
    \begin{equation}
    \text{S-Precision@K}=\frac{\left|\bigcup_{i=1}^k \mathcal{S}(d_i)\right|}{k}.
    \end{equation}
\end{itemize}

\vspace{-2mm}
\subsubsection{Social Welfare Metrics}
\label{sec:social_welfare_metric}
Diversity is not only a research problem of information retrieval in computer science.
Additionally, it has received lots of attention in other disciplines such as ecology and economics.
Recently, several works borrow the diversity notions from other fields for evaluating search and recommendation results.
We summarize these metrics as follows.

\begin{itemize}
     \item \textbf{SD Index} (short for \textbf{S}impson's \textbf{D}iversity Index).
    SD Index originated from paper~\cite{simpson1949measurement} for measuring the biodiversity in a habitat. 
    Regarding each subtopic (category) in recommendation as a kind of species in ecology, SD Index can be defined as the probability that two items selected randomly and independently without replacement belong to the same category. 
    Thus, a smaller SD Index indicates a higher diversity.
    We denote $n_S$ as the number of different subtopics, $l$ as the list of items under consideration (which can be a single recommendation list or the concatenation of multiple lists), and $c_{s_i}^l$ as the number of items covering the subtopic $s_i$ in the list $l$. 
    Then the SD Index over the list $l$ can be defined as follows:
    \begin{equation}
    \text{SD Index}=\frac{\sum_{i=1}^{n_S}\left[c_{s_i}^l\cdot\left(c_{s_i}^l-1\right)\right]}{|l|\left(|l|-1\right)}.
    \end{equation}
    
    To better demonstrate this, we assume $3$ subtopics in total.
    System $A$ recommends $10$ items and the number of items covering each subtopic is $8$, $1$, $1$.
    System $B$ also recommends $10$ items, while the number of items covering each subtopic is $4$, $3$, $3$.
    Then we can compute the SD Index of both systems and find out that the index value of system $A$ is larger than that of system $B$: $\frac{8\times 7+1\times 0+1\times 0}{10\times 9}>\frac{4\times 3+3\times 2+3\times 2}{10\times 9}$.
    This means that system $A$'s recommendation is less diverse than system $B$'s recommendation, which is aligned with our intuition.

    \item {\textbf{Gini Index}}.
    The Gini Index proposed by ~\citet{GiniIndex} is originally a measure of the distribution of income across a population. 
    A higher Gini Index indicates greater inequality, with high-income individuals receiving much larger percentages of the total income of the population.
    Recently, some researchers also adopt the Gini Index in the field of recommendation to measure the inequality among values of a frequency distribution, e.g., the number of occurrences (exposures) in the recommendation list.
    This measurement is generally at the system level by aggregating all the recommendation lists across all users, which can also indicate how diverse the system is in regard to all the items it can retrieve or recommend.
    Assuming the occurrence of the $i^{th}$ item is $e_i$, where $i=1, ...,  |\mathcal{D}|$, the Gini Index over all the items of the whole system is calculated as:
    \begin{equation}
    \setlength\abovedisplayskip{2pt}
    \setlength\belowdisplayskip{2pt}
        \text{Gini Index}=\frac{1}{2|\mathcal{D}|^2\overline{e}}\sum_{i=1}^{|\mathcal{D}|}\sum_{j=1}^{|\mathcal{D}|}|e_i-e_j|,
    \end{equation}
    where $\overline{e}=\frac{1}{|\mathcal{D}|}\sum_{i=1}^{|\mathcal{D}|}e_i$ is the mean occurrence of items. 
    Thus, a smaller Gini Index indicates a more fair distribution of the occurrences of items in the output results. This may indicate a higher diversity since different items have more equal opportunities to be exposed.

\end{itemize}

\vspace{-3mm}
\subsection{Relevance-aware Diversity Metrics}
Although diversity is an important property algorithm designers need to consider, relevance is still at the heart of the ranking problems in both search and recommendation.
Therefore, a metric that is solely concerned with diversity cannot adequately assess a system's effectiveness.
For instance, one can randomly select items from various topics to ensure that the ranking list performs exceptionally well on metrics such as S-Coverage and S-RR@100\%.
However, the overall relevance of the list obtained in such a way may be extremely low.
Therefore, if the algorithm designer aims to use the relevance-oblivious metrics to measure the diversity, she has to use other metrics (e.g., nDCG, rank-biased precision (RBP)~\cite{MoffatZ08RBP}) to measure the relevance.

In contrast to those relevance-obvious metrics on diversity, there exist relevance-aware metrics that attempt to incorporate both relevance and diversity into a single measurement, where almost all of them originated from research in search.
Several works conduct axiomatic analysis on the relevance constraints of diversity metrics~\cite{SIGIR2013:AmigoGV13, SIGIR2018:AmigoSA18}. 
Here, we highlight two of the most critical properties on the relevance in ranking, \textit{\textbf{priority}} and \textit{\textbf{heaviness}}, that the relevance-aware metrics in any ranking task must satisfy.
\begin{itemize}
    \item \textit{\textbf{Property 1: Priority}. Swapping items in concordance with their relevance scores should increase the overall score of the whole ranking.}
    
    We denote $o(d)$ as the relevance score of an item with respect to the query or user, $Q(\sigma)$ as the overall score of the ranking list $\sigma$, ``$\leftrightarrow$'' as swapping two items.
    Formally, the \textit{\textbf{priority}} property requires that: if $o(d_i)<o(d_j)$ and $\sigma^{-1}(d_i)<\sigma^{-1}(d_j)$, then $Q({\sigma}_{d_i\leftrightarrow{d_j}})>Q({\sigma})$.
    
    \item \textit{\textbf{Property 2: Heaviness}. Items with the same relevance score contribute more when at earlier positions in the ranking.}
    
    \new{Using the same notations, the \textit{\textbf{heaviness}} property requires that: if $o(d_i)=o(d_j)<o(d_{i'})=o(d_{j'})$, $\sigma^{-1}(d_{i'})-\sigma^{-1}(d_i)>0$, $\sigma^{-1}(d_j)-\sigma^{-1}(d_i)=\sigma^{-1}(d_{j'})-\sigma^{-1}(d_{i'})>0$, then $Q({\sigma}_{d_i\leftrightarrow{d_{i'}}})>Q({\sigma}_{d_j\leftrightarrow{d_{j'}}})$.}
\end{itemize}

It is easy to see that those widely used relevance metrics in information retrieval and recommendation, such as the nDCG, satisfy both of the two properties.
Now, we categorize the relevance-aware diversity metrics that satisfy the above properties into the following two categories.

\subsubsection{Novelty-based Metrics}
As a metric with a close connection to diversity, novelty was also studied in prior works. 
\citet{ClarkeKCVABM08alphaNDCG} point out the precise distinction between novelty and diversity in the field of information retrieval: \textit{novelty refers to the need to avoid redundancy, while diversity refers to the need to resolve ambiguity in the query}, which corresponds to \textit{intrinsic diversity} and \textit{extrinsic diversity}, respectively.
However, even with this difference, we still find out several works in the literature categorize the novelty-based metrics as part of the metrics in the diversity family, since novelty on topics and categories can also be regarded as an improvement on diversity.
We follow this paradigm and summarize the novelty-based metrics as follows.


\begin{itemize}
    \item \textbf{$\alpha$-nDCG@K} (short for Novelty-biased \textbf{N}ormalized \textbf{D}iscounted \textbf{C}umulative \textbf{G}ain@K). This metric is proposed by~\citet{ClarkeKCVABM08alphaNDCG}, using to balance the trade-off between retrieving both relevant and non-redundant items.
    This is also one of the earliest metrics aiming to combine the measurement of both relevance and diversity.
    Prior to this, most metrics in information retrieval and recommendation such as mean average precision (MAP) and nDCG assume that the relevance of each item can be judged in isolation, independently from other items, thus ignoring important factors such as redundancy between items.
    To address this, the authors present a framework for assessing diversity and novelty based on the cumulative gain.
    
    The key is to define the utility gain of adding the $k^{th}$ item ($d^k$) in the list to be considered in the left of all items ranked above position $k$. 
    The authors assume that subtopics are independent and equally likely to be relevant, and the assessment of positive relevance judgments of an item for a subtopic $o(d|s)$ involves an uncertainty that can be modeled with a fixed probability $\alpha$ of success in the judgment.

    
    We denote $c_{s_i}^{\sigma^{1:k}}$ as the number of items covering subtopic $s_i$ till position $k$ in the ranking list $\sigma$, then we can first formally define the $\alpha$-DCG@K over a ranking list $\sigma$ as below, \new{where $0<\alpha\leq1$~\cite{ClarkeKCVABM08alphaNDCG}}:
    \begin{align}
        \text{Gain}(d^k)&=\sum_{i=1}^{n_S}o(d^k|s_i)(1-\alpha)^{c_{s_i}^{\sigma^{1:k}}},\label{eq:Gain}\\
        \textbf{$\alpha$-DCG@K}&= \sum_{k=1}^{K}\frac{1}{\log(k+1)}\cdot \text{Gain}(d^k).
    \label{eq:alpha-DCG}
    \end{align}
    Analogous to the definition of nDCG@K, we can find an ``ideal'' ranking that maximizes $\alpha$-DCG@K, denoting as the $\alpha$-iDCG@K. 
    The ideal ranking computation is known to be an NP-Complete problem, pointed out by~\citet{Carterette09_NP_Diversity}. 
    The ratio of $\alpha$-DCG@K to $\alpha$-iDCG@K defines the $\alpha$-nDCG@K.

    \item \textbf{NRBP} (short for \textbf{N}ovelty- and \textbf{R}ank-\textbf{B}iased \textbf{P}recision).
    Following the paper~\cite{ClarkeKCVABM08alphaNDCG},~\citet{ClarkeKV09NRBP} propose another metric built upon the rank-biased precision (RBP), rather than the nDCG, with a very similar motivation and paradigm.
    Described by~\citet{MoffatZ08RBP}, the RBP model assumes that the user browses a list of items in sequential order and with a probability $\beta$ (i.e., $0<\beta<1$) to continue at each position.
    In other words, a user has a probability of $\beta^k$ to observe all the items till the $k^{th}$ position.
    Based on this idea and using the same notation as $\alpha$-nDCG@K in Eq.~\ref{eq:alpha-DCG}, the NRBP can be formally defined as follows:
    \begin{equation}
        \textbf{NRBP}=\frac{1-(1-\alpha)\beta}{n_S}\sum_{k=1}^{|\sigma|}\beta^{k-1}\sum_{i=1}^{n_S}o(d^k|s_i)(1-\alpha)^{c_{s_i}^{\sigma^{1:k}}}.
    \end{equation}
    Here, the normalization factor includes division by the number of subtopics, allowing us to average the measure across multiple queries with varying subtopic counts.
    It is also worth mentioning that in contrast to $\alpha$-nDCG@K which is typically presented at a particular browsing depth, NRBP is more of a summary metric that illustrates the diversity/novelty of the entire list.

    \item \textbf{nDCU@K} (short for \textbf{N}ormalized \textbf{D}iscounted \textbf{C}umulative \textbf{U}tility@K).
    The nDCU@K was proposed by~\citet{YangLLHKR07nDCU} around the same period when the $\alpha$-nDCG@K was proposed.
    It is also motivated by extending the original nDCG@K metric.
    Given a list of retrieved items based on a query, the authors define the utility of an item $d^k$ at the $k^{th}$ position as:
    \begin{equation}
        \text{Utility}(d^k)=\text{Gain}(d^k)-\text{Loss}(d^k),
    \label{eq:utility}
    \end{equation}
    where the $\text{Gain}(d^k)$ refers to the information users receive for observing $d^k$ depending on the relevance and novelty, while $\text{Loss}(d^k)$ denotes the time and energy spent in going through the item.
    There are various ways to define these two terms.
    \citet{YangLLHKR07nDCU} adopt a very similar formulation as that in Eq.~\ref{eq:Gain} to define the $\text{Gain}(d^k)$, and they define the $\text{Loss}(d^k)$ as a constant.
    Then the DCU@K can be defined as:
    \begin{equation}
        \textbf{DCU@K}= \sum_{k=1}^{K}\frac{1}{\log(k+1)}\cdot \text{Utility}(d^k).
    \end{equation}
    
    Analogous to the nDCG@K, the ratio of DCU@K to the ideal DCU@K defines the nDCU@K.

    Considering that nDCU@K is well-defined only for a single ranking list~\cite{YangLLHKR07nDCU}, \citet{YangL09EGU} extend the nDCU@K metric to make it be capable of measuring multiple ranking lists conditioned on different user browsing patterns.
    Specifically, they compute the mathematical expectation of Eq.~\ref{eq:utility} over $n$ ranking lists and $n$ user browsing models (one list and one browsing model for each user), where each browsing model corresponds to a list of $n$ different stop positions in the $n$ ranking lists. 
    Since this is just a slight modification of nDCU@K, we do not treat it as a separate metric in this survey.

\end{itemize}

\subsubsection{IA-family Metrics}
Sometimes, a simple combination of diversity and per-intent graded relevance is still not enough. 
When considering diversity, it is not always ideal to retrieve or recommend different items covering various topics without distinguishing which topic is more important.
Take the search problem as an example, ``intent'' is defined as the ``information needs'', which can also be referred as users' expectations on distributions of different ``subtopics'' in the search result with respect to a query.

The motivation begind intent-aware metrics can also be depicted by the following example described in paper~\cite{AgrawalGHI09Intent}.
Considering a query $q$ that is related to two subtopics $s_1$ and $s_2$, but is much more related to $s_2$ than $s_1$. 
Now we have two items $d_1$ and $d_2$, where $d_1$ rated 5 (out of 5) for $s_1$ but unrelated to $s_2$, $d_2$ rated 4 (out of 5) for $s_2$ but unrelated to $s_1$. 
Traditional relevance metrics tend to rank $d_1$ over $d_2$, but users may find $d_2$ is more related than $d_1$ on average.

As such,~\citet{AgrawalGHI09Intent} propose the family of intent-aware metrics for search results diversification.
Formally, given a distribution on the subtopics for a query and a relevance label on each item, they compute the outcome over a list $\sigma$ by applying the intent-aware scheme on a conventional relevance metric $M$ as follows:
\begin{align}
    \text{$M$-IA}(\bm{\sigma})=\sum_{i=1}^{n_S}p(s_i|q)\cdot \text{$M$}(\bm{\sigma}|s_i),
\end{align}
where $s$ is the subtopic, denoting the user intent.
$M$ is the traditional metric for measuring the ranking quality on relevance, such as nDCG, MRR, and MAP.
When computing the intent-dependent $M@K(\bm{\sigma}|s)$, they simply treat any item that does not match the subtopic $s$ as non-relevant items, then compute the same way as the original $M@K(\bm{\sigma})$.
Thus, the family of $M$-IA metrics takes into account the distributions of intents, and force a trade-off between adding items with higher relevance scores and those that cover intents with higher weights.

\subsubsection{D-family Metrics}

The intent-aware metrics are sub-optimal in that they are not guaranteed to range between 0 and 1: it is generally not possible for a single system output to be ideal for all intents at the same time. As such,~\citet{DBLP:conf/ntcir/SakaiCSRDL10,sakai2011evaluating} propose an alternative way to evaluate diversified search results, given intent probabilities and per-intent graded relevance assessments. While intent-aware measures combine multiple measure scores using the intent probabilities, the D-family combines per-intent relevance grades of each document using the intent probabilities. 

Given the intent probabilities $p(s_i|q)$ and per-intent graded relevance assessments, where $g(d^k|s_i)$ is the gain value for document at rank position $k$ for intent $s_i$, the global gain at rank position $k$ can be defined as:
\begin{equation}
    \text{GG($k$)} = \sum_{i = 1}^{n_S}p(s_i|q)\cdot g(d^k|s_i).
\end{equation}
For a launched query, an ideal ranking list can be obtained by sorting all documents by the global gain. 
Denoting the global gain of the document ranked at position $k$ in the retrieval list and the ideal ranking list, D-metrics can be computed by using these gain values instead of the traditional pre-set values.

Apart from the fact that D-metrics avoid the undernomalisation problem of intent-aware metrics by relying on a single “globally ideal” list, these two metric families are quite similar. However,~\citet{DBLP:conf/ntcir/SakaiCSRDL10,sakai2011evaluating} also propose a simple method to explicitly encourage high intent recall (S-Recall) in a search output within the D-metric framework. Then D\#-metrics are defined as:
\begin{equation}     
    \text{D\#-$M$@K} = \gamma\cdot \text{S-Recall@K} + (1-\gamma)\cdot \text{D-$M$@K}.
\end{equation}
where $\gamma$ is a parameter, $M$ is the traditional metric for measuring the ranking quality on relevance, such as nDCG, MRR, and MAP. 
Several other works such as~\cite{DBLP:journals/tkde/WangWDSZ18} propose further extensions of D/D\#-metrics by introducing intent hierarchies to model the relationships between intents, and present various weighing schemes.
In this survey, we do not classify them separately and still treat them as instances of the D-family metrics.

\section{Offline Approaches for Enhancing Diversity}
\label{sec:offline-approach}

\begin{figure*}[t]
    \centering
    \includegraphics[width=0.95\linewidth]{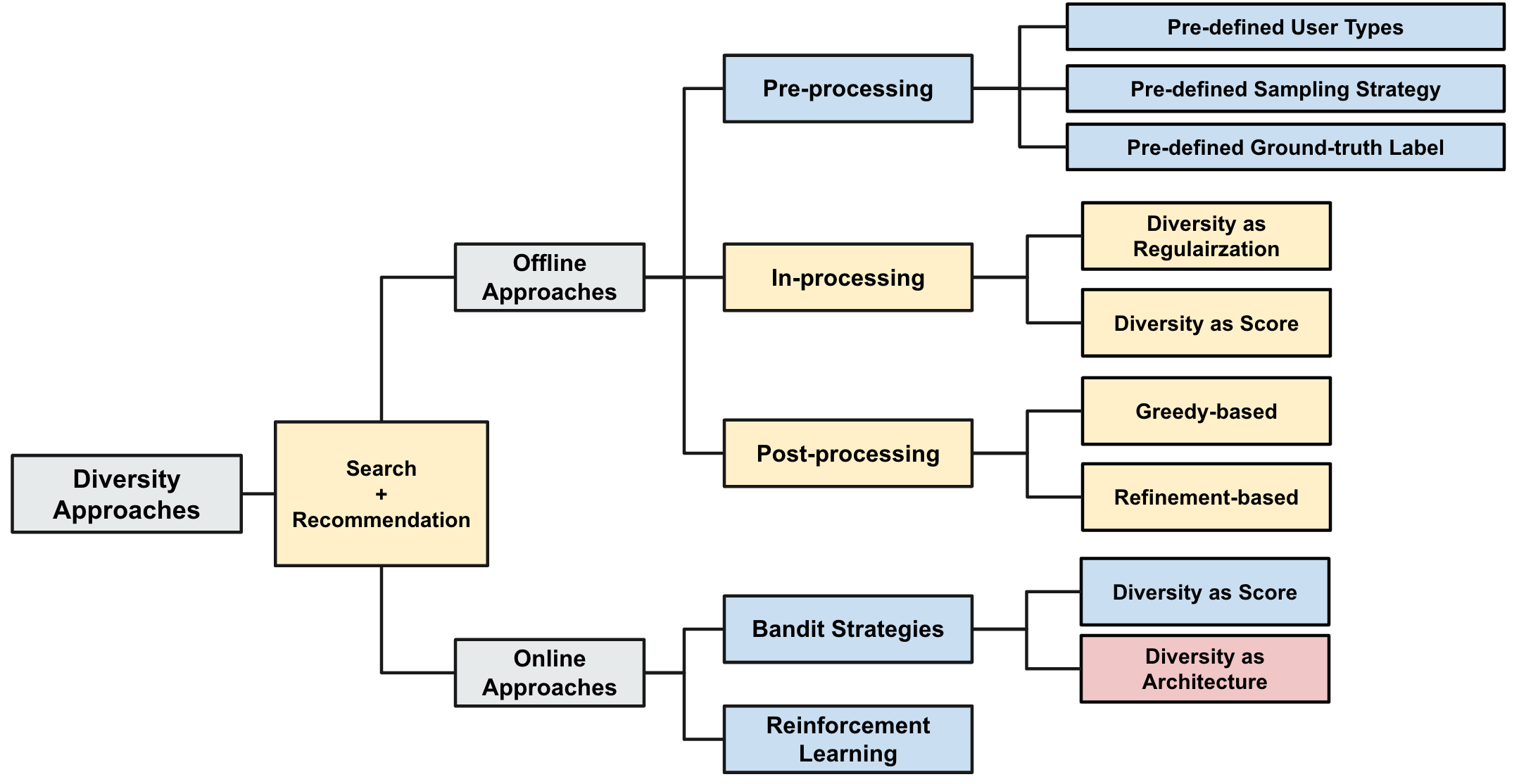}
    \caption{Diversity approaches in search and recommendation, from both offline and online perspectives. Approaches in blue boxes indicate that they are generally used in recommendation, the pink box indicates it is generally used in search, while those in yellow boxes are equally widely used in search and recommendation.}
    \label{fig:div_approach}
    \vspace{-2mm}
\end{figure*}

\begin{table*}[h]
\caption{Summary for the publications proposing or using different diversity approaches in search and recommendation.}
\label{tab:method}
\vspace{1mm}
\centering
\resizebox{\linewidth}{!}{
\begin{tabular}{l|l|l|l|l}

\Xhline{2\arrayrulewidth}
\multicolumn{4}{l|}{\textbf{Diversity Approaches}}                       &       \textbf{Related Work}                                                                                                                                                                                                                                                                           \\ \hline
\multirow{8}{*}{Offline Approaches} & \multirow{3}{*}{Pre-processing Methods} & \multicolumn{2}{l|}{Pre-defined User Types} & \cite{13KwonHH20} \\ 
\cline{3-5} 
& & \multicolumn{2}{l|}{Pre-defined Sampling Strategies} & \cite{ZhengGCJL21DGCN-pre} \\
\cline{3-5} 
& & \multicolumn{2}{l|}{Pre-defined Ground-truth Label} & \cite{DBLP:conf/recsys/PaudelHB17,ChengWMSX17Acc_Diverse} \\
\cline{2-5}
 & \multirow{2}{*}{In-processing Methods} & \multicolumn{2}{l|}{Diversity as Regularization} & \cite{CIKM2020:DBLP:conf/cikm/ChenRC0R20,CIKM2020:DBLP:conf/cikm/MaropakiCDN20,CIKM2020:DBLP:conf/cikm/ZhouAK20,DBLP:conf/recsys/Sanz-CruzadoC18,WasilewskiH16Diverse_LTR,DBLP:conf/sigir/BalloccuBFM22,chen2022dexdeepfm,chen2021multi} \\
 \cline{3-5}
 & & \multicolumn{2}{l|}{Diversity as Score} & \cite{CIKM2020:DBLP:conf/cikm/QinDW20,LiZZZL17DiverseMF,DBLP:conf/aaai/Yu22,han2019geographic}\\
 \cline{2-5}
  & \multirow{2}{*}{Post-processing Methods} & \multirow{2}{*}{Greedy-based} & MMR & \cite{CarbonellG98MMR,SantosMO10xquard,vargas2011intent, DBLP:journals/tkde/LiY13, DBLP:journals/tkde/0002LQXYZ16,raman2014understanding} \\
 \cline{4-5}
 & & & DPP & \cite{macchi1975DPP_origin,DBLP:conf/kdd/HuangWZX21,NIPS2018:DBLP:conf/nips/ChenZZ18,Wilhelm2018DPP2,Chen2017DPP1,GanLC20DPP4,DBLP:conf/sigir/GongZC0B0YQ22} \\
 \cline{3-5}
  & & \multicolumn{2}{l|}{Refinement-based} & \cite{li2020directional,DBLP:conf/recsys/TsukudaG19, DBLP:conf/www/ZieglerMKL05, YuLA09, DBLP:journals/tkde/AdomaviciusK12, DBLP:journals/tkde/YinHLZ13, DBLP:journals/tkde/LiY13, DBLP:journals/tkde/JiangDZNYW18,cai2016diversifying,zhang2016trip,raman2014understanding} \\
  \hline
\multirow{3}{*}{Online Approaches} & \multirow{2}{*}{Bandit Strategies} & \multicolumn{2}{l|}{Diversity as Score} & \cite{li2020cascading, DBLP:conf/aaai/DingLMCT21, QinCZ14CCB, WangWWH17BiUCB}\\
 \cline{3-5}
 & & \multicolumn{2}{l|}{Diversity as Architecture} & 
\cite{ParaparR21Bandit_Diverse_Arm, RadlinskiKJ08_RBA}\\
 \cline{2-5}
& \multicolumn{3}{l|}{Reinforcement Learning} & \cite{ZhengZZXY0L18DRN, DBLP:conf/wsdm/StamenkovicKAXK22} \\
\Xhline{2\arrayrulewidth}
\end{tabular}%
}
\vspace{-4mm}
\end{table*}

We intend to use a unified framework to categorize the approaches for enhancing diversity in both search and recommendation since there are lots of similarities in these two scenarios.
In this section, we focus on offline processes, where the methods do not need to care about users' real-time feedback based on the displayed results from the last round.
Based on when the approaches of diversity intervene relative to the training procedure, the offline diversity approaches can be divided into three categories: (\romannumeral1) \textbf{pre-processing}, (\romannumeral2) \textbf{in-processing}, and (\romannumeral3) \textbf{post-processing}.
Pre-processing methods are adopted prior to the model training process.
They typically pre-defined diversity-aware techniques with the expectation that these designs will result in a diverse output.
In-processing methods directly participate during the model training.
They guarantee a diversified outcome through learning matching scores for users and items where the diversity constraints or scores are added.
Post-processing methods are used after the model is well-trained.
They generally re-rank the final item lists to improve diversity.
All approaches and corresponding works are summarized in Table~\ref{tab:method}.

\vspace{-2mm}
\subsection{Pre-processing Methods}
Pre-processing methods intervene in the system before the model training. 
We review the following three sub-classes (\romannumeral1) \textit{pre-defined user types}, (\romannumeral2) \textit{pre-defined sampling strategies}, and (\romannumeral3) \textit{pre-defined ground-truth label}.

\subsubsection{Pre-defined User Types}
\citet{13KwonHH20} improve the diversity in recommendation through pre-defining user types.
They first interview fashion professionals and categorize the user purchase behavior into four types: (\romannumeral1) \textit{gift type}: purchasing items to give to others; (\romannumeral2) \textit{coordinator type}: purchasing items associated with previous purchases; (\romannumeral3) \textit{carry-over type}: purchasing items similar to existing purchases; and (\romannumeral4) \textit{trend-setter type}: purchasing items affected by the trends of other people’s purchases.
For each type, they use a specific algorithm to recommend top-5 items to each user.
Since each user may have multiple purchase behaviors, they adopt a hybrid strategy to simply combine the recommendation lists of four algorithms to generate up to 20 item candidates for each user.
The C-Coverage improves from 24\% (using the CF algorithm) to 78\%, coupled with a 3.2\% increase in the purchasing rate and a \$13 increase in the average purchase amount per customer in the experimental group. 

\subsubsection{Pre-defined Sampling Strategies} 
Since the interactions between users and items can be naturally represented as a graph, a number of graph-based recommendation algorithms are widely used nowadays, such as the Graph Neural Network (GNN)-based models.
These models represent the users and items as nodes in a graph, followed by learning their embeddings through message passing: at each layer, they aggregate the neighbor's information for each target node.
Due to their ability to capture higher-order connectivity between user nodes and item nodes, GNN-based methods can generally achieve state-of-the-art accuracy and relevance.

\begin{figure}[t]
    \centering
    \includegraphics[width=1.0\linewidth]{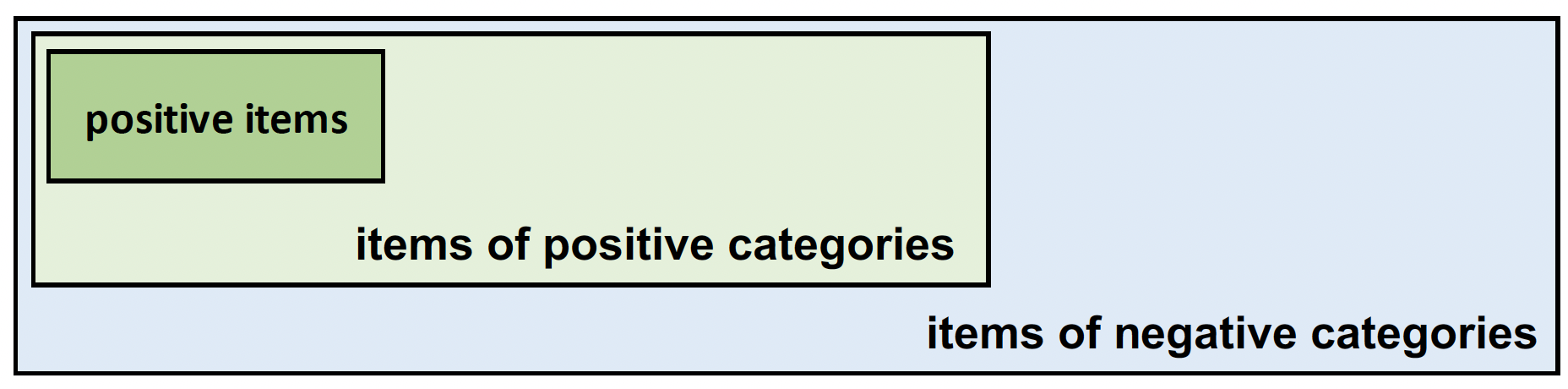}
    \caption{An illustration of the category-boosted negative sampling. Negative items are sampled from outside positive items. The strategy boosts the probability of sampling from items of positive categories (the light green area). The figure is borrowed from paper~\cite{ZhengGCJL21DGCN-pre}.}
    \label{fig:pre_neg_sample}
    \vspace{-4mm}
\end{figure}

Since higher-order neighbors of a user tend to cover more diverse items, GNN-based approaches have the potential to improve recommendation diversity as a byproduct.
Without specific design, those items from the popular categories tend to be learned more often, because they take up the majority of the edges on the graph.
To address this,~\citet{ZhengGCJL21DGCN-pre} propose two pre-defined sampling strategies for two processes in the model.
The first strategy aims to re-balance the \textit{neighborhood sampling} process in the message passing to increase the selecting probabilities for those items from the less popular categories and decrease those from the popular categories.
In such a way, those less popular items can still have a chance to be sampled and well-learned.
The second strategy affects the \textit{negative sampling} process.
In contrast to random negative sampling in paper~\cite{DBLP:conf/uai/RendleFGS09}, they propose to select similar (from the same category) but negative items with an increased probability, so that less similar (not from the same category) items are not pushed too far away from the user in the embedding space.
As a result, items from different categories are likely to appear in the recommendation list for each user, thus enhancing the individual-level diversity.
An illustration of this category-boosted negative sampling is shown in Fig.~\ref{fig:pre_neg_sample}. 

\subsubsection{Pre-defined Ground-truth Label} 
\citet{ChengWMSX17Acc_Diverse} construct ground truth labels via diversity constraints to directly idealize the optimization target. 
Employing supervised learning, each user becomes a training instance with a heuristically chosen subset of relevant, diverse items as their ground-truth label.
Their two-step labelling method involves: (\romannumeral1) filtering high-rated items into a candidate set $\mathcal{C}_u$, and (\romannumeral2) selecting items from $\mathcal{C}_u$ to maximize the relevance-diversity balance. 
An item $d$ qualifies as high-rated for user $u$ if $o(d|u)\geq\gamma\cdot\overline{o}(\cdot|u)$, where $o(d|u)$ denotes the observed score of $d$ by $u$, $\overline{o}(\cdot|u)$ is the average score across all items by $u$, and $\gamma$ is a trade-off parameter.

All selected items in step (\romannumeral1) form the set of candidates $\mathcal{C}_u$.
In step (\romannumeral2), they select a subset $\mathcal{Y}_u$  from $\mathcal{C}_u$ (i.e., $\mathcal{Y}_u\subseteq\mathcal{C}_u$, $|\mathcal{Y}_u|=K$) as the ground-truth label for user $u$ by balancing the trade-off between relevance and diversity, using a metric similar to the F-measure~\cite{Baeza-YatesR99_Fmeasure}.
Specifically, the selected $\mathcal{Y}_u$ aims to maximize the following equation:
\begin{align}
\max\limits_{\mathcal{Y}_u}\quad&\frac{2\cdot f(\mathcal{Y}_u)\cdot g(\mathcal{Y}_u)}{f(\mathcal{Y}_u)+g(\mathcal{Y}_u)},\\
&\text{s.t.,} \quad \mathcal{Y}_u\subseteq\mathcal{C}_u, |\mathcal{Y}_u|=K.
\end{align}
Here, $f(\mathcal{Y}_u)$ and $g(\mathcal{Y}_u)$ represent the measurement for relevance and diversity over the whole set $\mathcal{Y}_u$, respectively.
For the relevance, denoting the set of items rated by $u$ as $\mathcal{D}_u$, \citet{ChengWMSX17Acc_Diverse} define $f(\cdot)$ as a pair-wise comparison on the ratings between the items in $\mathcal{Y}_u$ and $\mathcal{D}_u\backslash\mathcal{Y}_u$ as follows:
\begin{small}
\begin{align}
f(\mathcal{Y}_u)=\frac{\sum\limits_{d_i\in\mathcal{Y}_u}\sum\limits_{d_j\in\mathcal{D}_u\backslash\mathcal{Y}_u}\text{compare}\bigg(o(d_i|u)-o(d_j|u)\bigg)}{|\mathcal{Y}_u|\cdot |\mathcal{D}_u\backslash\mathcal{Y}_u|},
\end{align}
\end{small}
where $\text{compare}(x)$ equals 1 if $x>0$; else, equals -1.
For the diversity measurement $g(\cdot)$, the authors define it as the ILAD as described in Eq.~\ref{eq:ilad}.
Afterward, the obtained ground-truth label $\mathcal{Y}_u$ for user $u$ can guide the model training.

\vspace{-2mm}
\subsection{In-processing Methods}
In-processing methods act during the model training process. 
We categorize them into the following two sub-classes: (\romannumeral1) \textit{diversity as regularization} and (\romannumeral2) \textit{diversity as score}.

\subsubsection{Diversity as Regularization}
Since relevance is the primary goal of search and recommendation systems, the most intuitive way to enhance diversity through an in-processing way is to treat diversity as a regularization on the loss function to guide the training.
\citet{WasilewskiH16Diverse_LTR} first propose a prototype to constrain the relevance loss with a trade-off parameter $\lambda$:
\begin{equation}
    \min\limits_{\Theta} \mathcal{L}_{\text{rel}}(\Theta) + \lambda\cdot \mathcal{L}_{\text{div}}(\Theta),
\end{equation}
where $\Theta$ is the learnable embeddings, $\mathcal{L}_{\text{rel}}(\cdot)$ and $\mathcal{L}_{\text{div}}(\cdot)$ refer to the relevance loss and diversity regularization which can be both self-defined.
For instance,~\citet{WasilewskiH16Diverse_LTR} define the relevance loss as the pair-wise ranking loss~\cite{DBLP:conf/uai/RendleFGS09}, and the diversity loss as the negative of the intra-list average distance (ILAD)~\cite{DBLP:conf/recsys/ZhangH08}.

Several later works follow this line of research, such as paper~\cite{CIKM2020:DBLP:conf/cikm/ChenRC0R20}.
Rather than only modeling the dissimilarities among items for defining the diversity, the authors take the user intents into consideration.
Here, the user intents can be comprehended as the user's interest in different subtopics (categories).
Specifically, the authors define the diversity of a recommendation list $\sigma_u$ to user $u$ as the probability of each subtopic $s_i$ having at least one relevant item in $\sigma_u$, then the regularization can be formulated as:
\begin{equation}
    \mathcal{L}^{\sigma_u}_{\text{div}}=-\sum\limits_{i=1}^{n_S}p(s_i|u)\cdot\bigg(1-\prod\limits_{d\in\sigma_u}[1-p\left(d|s_i\right)]\bigg).
~\label{eq:intent-div-loss}
\end{equation}
Here, $p(s_i|u)$ represents $u$'s interest in subtopic $s_i$, and $p\left(d|s_i\right)$ refers to the relatedness of $d$ to the subtopic $s_i$.
Both terms can be computed through a softmax function using the embeddings of users, items, and subtopics.

In addition to merely focusing on the diversification of retrieval results, some researchers also care about how to generate diverse explanations for the output results. 
For instance,~\citet{DBLP:conf/sigir/BalloccuBFM22} conceptualize, assess, and operationalize three novel properties (linking interaction recency, shared entity popularity, and explanation type diversity) to monitor explanation quality at the user level in recommendation, and propose re-ranking approaches able to optimize for these properties. 
They optimize these three indicators for measuring the quality of explanations as a regularization term in the re-rank stage. 
Here we classify this as an in-processing method.

\vspace{-2mm}
\subsubsection{Diversity as Score}
\label{sec:div_as_score}
Another widely adopted in-processing method for diversity is to treat diversity as a score during ranking.
As such, the score of an item is composed of two parts: one from the perspective of relevance, and the other from the perspective of diversity.
The most significant difference between these two types of scores is that the relevance score typically assumes the independence of items in a list, but the diversity score is highly dependent on the other items. 

Following this line,~\citet{LiZZZL17DiverseMF} propose one of the earliest methods of diversified recommendation.
They focus on a sequential recommendation process, where the model recommends one item at a time to form the entire recommendation list.
They define the score of an item at the current position as the sum of a relevance part and a discounted subtopic diversification part.
In detail, for user $u$, given an un-selected $k^{th}$ item $d^k$ and a list of selected $k-1$ items $\sigma^{1:k-1}_u$ \new{(i.e., the list of first $k-1$ items in the ranking list $\sigma_u$)}, they define the score of $d^k$ as: 
\begin{equation}
    o(d^k|u)=o^{\text{rel}}(d^k|u)+\lambda\cdot o^{\text{div}}(d^k|\sigma^{1:k-1}_u, u),
\label{eq:Diverse_score}
\end{equation}
where $o^{\text{rel}}(d^k)$ and $o^{\text{div}}(d^k)$ denote the score of $d^k$ from the view of relevance and diversity, respectively. 
Specifically, the authors define the relevance score as the inner product between the user and item embedding: $o^{\text{rel}}(d^k|u)=\Theta_u\cdot \Theta_{d^k}^\intercal$.
They define the diversity score as discounted subtopic's contribution, which reduces exponentially as the number of items covering that subtopic increases in the entire list:
\begin{align}
    o^{\text{div}}(d^k|\sigma^{1:k-1}_u,u)&=\sum\limits_{i=1}^{n_S}\beta^{c_{s_i}^{\sigma_u^{1:k}}}\cdot \Theta_u\cdot \Theta_{s_i}^\intercal,\\ \sigma_u^{1:k}&=\sigma_u^{1:k-1}.\text{append}(d^k).
\end{align}
Here $\beta$ is the decay factor (i.e., 0$<\beta<$1), $\Theta_u$ is the embedding of user $u$, $\Theta_{s_i}$ is the embedding of subtopic $s_i$, $c_{s_i}^{\sigma_u^{1:k}}$ denotes the number of items covering subtopic $s_i$ in $\sigma^{1:k}_u$.
As such, for each user, the model greedily selects an un-selected item to maximize the score in Eq.~\ref{eq:Diverse_score} at each position to form the final recommendation list.

To train embedding $\Theta$,~\citet{LiZZZL17DiverseMF} assume that the ideal recommendation lists for some sample users are available.
Then, the learning process aims to penalize those generated recommendations which do not respect the sequence in ideal lists.
Taking an individual user $u$ as an example, the loss function on a pair of sampled items $(d_i, d_j)$ can be formulated through a binary cross-entropy loss:
\begin{equation}
\mathcal{L}_u(d_i, d_j)=-y_{ij}\cdot\log(p_{ij})-(1-y_{ij})\cdot\log(1-p_{ij}),
\end{equation}
where $y_{ij}=1$ if $d_i$ is ranked above $d_j$ in the ideal ranking list of $u$, $p_{ij}$ refers to the probability of ranking $d_i$ above $d_j$ in the current model, which is computed as $p_{ij}=\text{sigmoid}(o(d_i|u)-o(d_j|u))$.

Other works also follow a similar idea to treat diversity as a score.
For instance,~\citet{DBLP:conf/aaai/Yu22} presents a novel framework for search result diversification based on the score-and-sort method using direct metric optimization. 
They express each item's diversity score specifically, which determines its rank position based on a probability distribution.

\vspace{-3mm}
\subsection{Post-processing Methods}
Earliest diversity approaches follow the re-ranking paradigm: they achieve diversity after the training procedure by re-ranking the list based on both relevance scores and diversity metrics.
Due to the separation of model training and diversified ranking, these approaches are regarded as post-processing, which can be applied to any recommendation models as a consecutive layer with excellent scalability.
Based on how the diversified list is generated, we categorize them as (\romannumeral1) \textit{greedy-based} and  (\romannumeral2) \textit{refinement-based}.



\vspace{-2mm}
\subsubsection{Greedy-based Methods}
As the name suggests, greedy selection methods iteratively select the item that maximizes a joint measure of relevance and diversity to each position, and finally provide an output ranking list.
Two of the most representative post-processing methods in this category are (\romannumeral1) \textit{Maximal Marginal Relevance (MMR)} and (\romannumeral2) \textit{Determinantal Point Process (DDP)}.

\vspace{-2mm}
\paragraph{\textbf{MMR}}
Maximal Marginal Relevance (MMR)~\cite{CarbonellG98MMR} is the most pioneering diversity approach in this category.
\citet{CarbonellG98MMR} propose ``marginal relevance'' as a linear combination of the relevance and diversity of each item, in response to the fact that user needs include not only \textit{relevance} but also \textit{novelty} and \textit{diversity}.
In particular, an item has high marginal relevance if it is both relevant to the user and has low similarity to previously selected items.

Based on this protocol, MMR greedily selects the item that can maximize the marginal relevance to form the final ranking list. 
We can formulate the process of MMR selecting the $k^{th}$ item for user $u$ as follows:
\begin{small}
\begin{equation}    d^k=\max\limits_{d\in(\mathcal{D}\backslash\mathcal{D}_u)\backslash\text{set}(\sigma^{1:k-1}_u)}[o^{\text{rel}}(d|u)+\lambda\cdot o^{\text{div}}(d|\sigma^{1:k-1}_u, u)],
\end{equation}
\end{small}
\new{where $\text{set}(\sigma^{1:k-1}_u)$ is the set of items composed by the first $k-1$ items in the ranking list $\sigma_u$.}

Different researchers adopt similar ways to model $o^{\text{rel}}(\cdot)$ as the inner product between the embeddings, while adopting different ways to model the $o^{\text{div}}(\cdot)$.
\citet{CarbonellG98MMR} define the diversity term as:
\begin{equation}
    o^\text{div}(d|\text{set}(\sigma^{1:k-1}_u), u)=-\max\limits_{d_j\in\text{set}(\sigma^{1:k-1}_u)}\text{sim}(d, d_j).
\end{equation}

One may observe that the recommendation generation process of MMR is quite similar to that of paper~\cite{LiZZZL17DiverseMF}, which falls under the category ``In-processing - Diversity as Score''.
The difference between these two works is as follows.
MMR is a post-processing method that performs a greedy selection based on already learned model embeddings.
In contrast, paper~\cite{LiZZZL17DiverseMF} adopts an in-processing method, whose greedy selection is not based on well-trained embeddings.
In other words, the diversification of MMR is added after model training, while the diversification in paper~\cite{LiZZZL17DiverseMF} is added during the model training procedure.
This is also the primary distinction between any post-processing and in-processing approaches.

\vspace{-2mm}
\paragraph{\textbf{DPP}}
Determinantal Point Process (DPP) is one of the cutting-edge post-processing methods for diversity enhancement in search and recommendation.
First introduced by~\citet{macchi1975DPP_origin} with the name ``fermion process'', DPP was originally used to precisely describe the repulsion and diversity phenomenon for fermion systems in thermal equilibrium.
Recently, it has been applied in search and recommendation for enhancing diversity~\cite{Chen2017DPP1}.

Prior to the application of DPP in the recommendation, most diversity approaches, such as the basic version of MMR~\cite{CarbonellG98MMR}, compute the similarity between items in a pair-wise way and avoid recommending redundant items to improve diversity.
However, these methods are sub-optimal since the pair-wise dissimilarities may not capture complex similarity relationships within the whole list, also the relevance and diversity are captured separately~\cite{Chen2017DPP1}.
Thanks to DPP's outstanding ability to capture the global correlations among data with an elegant probabilistic model~\cite{KuleszaT12DPP4ML}, DPP-based methods directly model the dissimilarities among items in a set-wise way using a unified model.

The idea of DPP can be demonstrated as follows.
A point process $\mathcal{P}$ on a set $\mathcal{D}$ (e.g.,
a set of $|\mathcal{D}|$ items) is a probability distribution on the powerset of $\mathcal{D}$ (the set of all subsets
of $\mathcal{D}$). 
That is, $\forall \mathcal{C}\subseteq\mathcal{D}$, $\mathcal{P}$ assigns some probability $p(\mathcal{C})$, and $\sum_{\mathcal{C}\subseteq\mathcal{D}}p(\mathcal{C})=1$.
Although a DPP defines a probability distribution over an exponential number of sets, it can be compactly parameterized by a single positive semi-definite (PSD) matrix $\mathbf{L}\in\mathbb{R}^{|\mathcal{D}|\times |\mathcal{D}|}$~\cite{BorodinDPP}.
The probability of a subset $\mathcal{C}$ represented by a DPP can be written as:
\begin{align}
    p(\mathcal{C})\propto\text{det}(\mathbf{L}_{\mathcal{C}}).
\label{eq:DPP_def}
\end{align}
where $\mathbf{L}_{\mathcal{C}}=[\mathbf{L}_{ij}]_{d_i, d_j\in\mathcal{C}}$.
However, it is still unclear how the determinant unifies the relevance and diversity.
To show this, we offer two views.
The first comprehension is based on the geometric meaning of determinant.
Since $\mathbf{L}$ is PSD, we can find a matrix $\mathbf{B}$ such that $\mathbf{L}=\mathbf{B}^\intercal\mathbf{B}$.
Then we have $\text{det}(\mathbf{L}_{\mathcal{C}})=\text{Vol}^2(\{\mathbf{B}_i\}_{i\in\mathcal{C}})$, which can be represented as the squared volume of the parallelepiped spanned by the columns of $\mathbf{B}$ corresponding to elements in $\mathcal{C}$.
The volume here is determined by two factors: the magnitude of column vectors and the orthogonality among them.
If we treat the columns of the matrix $\mathbf{B}$ as item embeddings, then it is clear to see that a larger magnitude of the vectors (\underline{\textit{higher relevance}}) and stronger orthogonality among them (\underline{\textit{higher dissimilarity}}) lead to a higher volume (\underline{\textit{higher determinant}}).
Thus, the determinant of a matrix unifies both relevance and diversity.
The second comprehension is based on a simple case where the matrix is $2\times 2$:
$\mathbf{L}_{\mathcal{C}}=\left [\begin{array}{cccc}
l_{11} & l_{21} \\
l_{12} & l_{22}
\end{array}\right]$,
whose determinant is $\text{det}(\mathbf{L}_{\mathcal{C}})=l_{11}\cdot l_{22}- l_{21}\cdot l_{12}$. 
Assuming the diagonal entries indicate the relevance of items and the off-diagonal entries indicate the similarity among items, then the determinant can be represented as relevance minus similarity.
Although this comprehension is for a 2-dimensional case, a similar intuition holds for higher dimensions.
Based on the above comprehension and intuition, \citet{Chen2017DPP1} construct user-specific $\mathbf{L}$ as $\mathbf{L}=\text{diag}(\bm{\xi})\cdot \mathbf{S}\cdot\text{diag}(\bm{\xi})$,
where $\text{diag}(\bm{\xi})$ is a diagonal matrix whose diagonal entries are the relevance scores between items to the user, and the $(i, j)$-th element of $\mathbf{S}$ is the similarity score between the $i^{th}$ item and the $j^{th}$ item.
Thus, Eq.~\ref{eq:DPP_def} can also be written as:
\begin{align}
    p(\mathcal{C})&\propto\text{det}(\mathbf{L}_{\mathcal{C}})=\left(\prod\limits_{i\in\mathcal{C}}\xi_i^2\right)\cdot\text{det}({\mathbf{S}_{\mathcal{C}}}).
\end{align}

Finally, to obtain the diversified top-$K$ recommendation for user $u$ given the whole item set $\mathcal{D}$, we first construct a user-specific matrix $\mathbf{L}$ as aforementioned.
Then the task can be formulated as follows:
\begin{equation}
\text{set}(\sigma_u)=\argmax_{\mathcal{C}\subseteq (\mathcal{D}\backslash\mathcal{D}_u), |\mathcal{C}|=K}\text{det}(\mathbf{L}_{\mathcal{C}}).
\label{eq:DPP_task}
\end{equation}
However, directly solving this task is expensive.
Approximate solutions to Eq.~\ref{eq:DPP_task} can be obtained by several algorithms, among which the greedy solution was previously considered the fastest one. 
Initializing $\sigma_u$ as empty, an item $d$ that maximizes the following equation is added to $\sigma_u$ iteratively.
Specifically, the selection of the $k^{th}$ item for user $u$ can be described as follows:
\begin{equation}
\resizebox{0.89\columnwidth}{!}{%
$
d^k=\argmax\limits_{d\in (\mathcal{D}\backslash\mathcal{D}_u)\backslash{\text{set}(\sigma_u^{1:k-1})}}\left(\text{det}(\mathbf{L}_{{\text{set}(\sigma_u^{1:k-1}})\cup\{d\}}) - \text{det}(\mathbf{L}_{{\text{set}(\sigma_u^{1:k-1})}})\right).
$
}
\end{equation}

Based on the strength of DPP,~\citet{DBLP:conf/sigir/GongZC0B0YQ22} propose a diversity-aware Web APIs recommendation methodology for choosing diverse and suitable APIs for mashup creation. 
The APIs recommendation issue for mashup creation is specifically treated as a graph search problem that seeks the smallest group of Steiner trees in an API correlation graph. 
Their method innovatively employs determinantal point processes to diversify the recommended results.

\subsubsection{Refinement-based Methods}
Unlike greedy-based methods that iteratively select items to form the entire ranking list, refinement-based methods adjust positions or replace items in existing ranking lists. Typically, using refinement-based methods, items are initially ranked using relevance metrics and subsequently refined by introducing diversity metrics.

Several earlier works follow this line of approach.
For instance,~\citet{DBLP:conf/www/ZieglerMKL05} construct two ranking lists for retrieving the diversified top-K items: $\sigma_{rel}$ and $\sigma_{div}$, where the first list is constructed merely based on the relevance score, while the second one is constructed based on the diversity score of each item in the whole candidate sets.
Both lists rank the items in descending order based on scores.
For achieving a single diversified ranking list, the authors merge the two lists using a scaling factor to trade-off how much to rely on the rankings in $\sigma_{rel}$ or $\sigma_{div}$.
A similar strategy is used by~\citet{YuLA09}. 
Starting from one ranking list with the $K$ highest scoring items, the authors swap the item that contributes the least to the diversity of the entire set with the next highest scoring item from the remaining. 
They set a threshold for the relevance when replacing the items in order to avoid a dramatic drop in the overall relevance.



\vspace{-3mm}
\section{Online Approaches for Diversity}
\label{sec:online-approach}
So far, we have reviewed the offline approaches for enhancing diversity in search and recommendation. 
These methods generally train the model in an offline manner using the existing data with ground-truth labels. 
However, in some situations, these labeled data are insufficient or unavailable, especially in the recommendation scenario.
For instance, one of the most well-known challenges is the ``cold-start'' problem where new users join the system.
To resolve these problems, one effective way is to use online approaches where the systems first display item lists to users, gather user feedback, and then update the model for the next turn.
Based on whether the user preference change, we further divide them as (\romannumeral1) \textit{bandit strategies} (i.e., invariant user preference) and (\romannumeral2) general \textit{reinforcement learning} (i.e., dynamic user preference). 
In this section, we review how to achieve diversity in these approaches.

\vspace{-3mm}
\subsection{Bandit Strategies}
As one of the simplest examples of reinforcement learning (RL), the bandit problem was first introduced by~\citet{Thompson33_bandit} in 1933. 
The most classical bandit problem is known as the multi-armed bandit (MAB), whose name comes from imagining a gambler at a row of slot machines, who has to decide how to play these machines to gain as much money as possible in a time horizon~\cite{Weber92_Gittins}.
A bandit problem can be generally defined as a sequential game between an agent and an environment~\cite{lattimore20_bandit}.
The game is played over $T$ rounds (i.e., the time horizon), while in each round $t\in[T]$, the agent first chooses an action $A_t$ from a given set $\mathcal{A}$, and the environment then reveals a reward $r_t\in R$. 
The goal of the agent is to maximize the $T$-step cumulative reward or, equivalently, minimize the $T$-step cumulative regret.
Here, the cumulative regret is defined as the expected difference between the reward sum associated with an optimal strategy and the sum of the collected rewards $\rho=T\cdot\mu^*-\sum_{t=1}^T r_t$, where $\mu^*$ is the maximal reward mean associated with the optimal strategy.

It is intuitive to model online search and recommendation as an MAB, where the algorithm is the agent, items are arms, displaying an item is selecting the corresponding arm, and user feedback is the reward. 
However, MAB does not use state information, or context (i.e., user and item features), which can limit performance, particularly in recommendation where personalization is key. 
To address this, most works use an MAB extension called Contextual MAB (CMAB) for online search and recommendation problems. 
Abundant work shows that CMAB typically outperforms MAB in the relevance of output lists.

Due to the simplicity of implementation and capability of making real-time decisions, recent research also aims to incorporate diversity in bandit algorithms for search and recommendation.
There are generally two ways to enhance diversity in these methods: either to treat diversity as part of the scores of each arm or to design a different bandit architecture that can lead to a diversified result.
We review both of these two ideas in the following paragraphs.

\vspace{-2mm}
\subsubsection{Diversity as Score}
Most works interpret diversity as part of the score on each arm (item) in the bandit algorithms for search and recommendation. 
For instance,~\citet{li2020cascading} formulate the diversified retrieval of top-$K$ items as a bandit problem with cascading user behavior, where a user browses the displayed list from top to bottom, clicks the first attractive item, and stops browsing the rest.
If the user clicks an item, the reward is 1, otherwise 0.
Then the objective is to minimize the following $T$-step cumulative regret:
\begin{equation}
    R(T)=\sum_{t=1}^{T}\mathbb{E}[r(\sigma^*, \alpha_t)-r(\sigma^t, \alpha_t)].
\end{equation}
Here, $r(\cdot)$ is the binary reward from the user feedback.
$\sigma^t$ is the displayed ranking list at time step $t$, while $\sigma^*$ is the optimal ranking list, with constraint that $|\sigma^t|=|\sigma^*|=K$.
$\alpha_t$ is a vector of length $K$, indicating the \textit{attraction} of each arm (item) in the ranking list at time step $t$, where is how diversity comes in.
Specifically, the authors define the \textit{attraction} as a combination of relevance and diversity, following a very similar way to Eq.~\ref{eq:Diverse_score} in Section~\ref{sec:div_as_score}.
Again, all the definitions of diversity are applicable, while both paper~\cite{li2020cascading} and~\cite{DBLP:conf/aaai/DingLMCT21} use the gain on coverage of subtopics (S-Coverage) of adding item $d_i$ as the attraction score of which from the diversity component. 
Several other works choose different ways to define the diversity score.
For instance,~\citet{QinCZ14CCB} use the entropy regularizer, while~\citet{WangWWH17BiUCB} propose three separate solutions, borrowing from MMR~\cite{CarbonellG98MMR}, entropy regularizer~\cite{QinCZ14CCB}, and temporal user-based switching~\cite{LathiaHCA10Time}.

\vspace{-2mm}
\subsubsection{Diversity as Architecture}
Rather than merely treating diversity as part of the score,~\citet{ParaparR21Bandit_Diverse_Arm} design a different bandit architecture for enhancing diversity.
Different from prior works that interpret each arm as an individual item, the authors first make each arm represent a unique item category, and further consider retrieving different items under each category. 
Such a two-stage design can not only guarantee the items are diverse (i.e., satisfy the distance-based metrics), but also guarantee different categories are covered as much as possible (i.e., satisfy coverage-based metrics).
In such a way, the algorithm can be efficiently used to construct user profiles with diverse preference elicitation.

All the works above lie in the recommendation scenario, where the personalization is at the core.
However, the output of a conventional web search is typically static, so it is more concerned with satisfying a population of users as opposed to each individual.
Following this line,~\citet{RadlinskiKJ08_RBA} propose to learn diverse rankings in web search systems through MAB. 
Their proposed approach, Ranked Bandits Algorithm (RBA), runs an MAB instance $\text{MAB}_i$ for each rank $i$ (i.e., $1\leq i\leq K$), where the arm of each MAB indicates a unique item.
When user $u_t$ arrives at time $t$, each $\text{MAB}_i$ sequentially and independently decides which item to select at the rank $i$ for displaying to $u_t$.
Assuming $u_t$ follows a cascading browsing behavior (i.e., click at most one relevant item in the list), if $u_t$ clicks on an item actually selected by an MAB instance, the reward for the arm corresponding to that item for the MAB at that rank is 1.
The reward for the arms corresponding to all other items is 0.
As such, each MAB can update the value of each item iteratively through multiple rounds.
Although RBA shows effectiveness both empirically and theoretically, it is worth noting that it is hard to be extended to non-binary payoffs.

\vspace{-2mm}
\subsection{Reinforcement Learning}
Although bandit strategies show efficieny and effectiveness in online search and recommendation, there exist several obvious limitation in them. 
Firstly, bandit algorithms has only one state with several actions that lead back to the same state.
In other words, they assume that user preference will always remain the same, which does not hold in most real-world scenarios.
Secondly, bandit algorithms only care about the immediate reward, while the long-term reward is still significant to real-world users.
To address these, most research naturally brings in the reinforcement learning (RL) framework to model the problem, where the state can be affected by the action of agents and the long-term reward is also captured during recommendation. 

In the RL setting, diversity has been promoted by employing efficient exploration-exploitation strategies. 
\citet{ZhengZZXY0L18DRN} first use a Deep Q-Learning Network (DQN)~\cite{MnihKSRVBGRFOPB15DQN} to capture the long-term award of users' actions.
As for the diversity, they adopt a Dueling Bandit Gradient Descent (DBGD)~\cite{GrotovR16tutorial, HofmannSWR13fastIR, YueJ09dueling} algorithm to do exploration in the DQN framework.
Specifically, during their exploration strategy, the agent aims to generate a recommendation list $\sigma$ using the current network $Q$ and another list $\sigma'$ using an explore network $Q'$, which shares the same architecture as $Q$ with a small disturbation added on the paratemers of $Q$.
Then the authors conduct a probabilistic interleave~\cite{GrotovR16tutorial} to generate the merged recommendation list based on $L$ and $L'$ for obtaining a diversified ranking list.
Other researchers such as~\citet{DBLP:conf/wsdm/StamenkovicKAXK22} first define and presente the next item recommendation objective as a Multi-objective MDP problem. 
Thereafter, they propose a Scalarized Multi-objective Reinforcement Learning (SMORL) model, which works as a regularizer, incorporating desired properties into the recommendation model to balance the relevance, diversity, and novelty of recommendation.
\vspace{-2mm}
\section{\new{Applicability of Diversity Metrics and Approaches}}
\label{sec:applicability}

\new{We now delve into the applicability of diversity metrics and approaches across various recommendation models.}

\new{Diversity metrics are largely model-agnostic, offering a versatile toolkit for evaluating across a wide array of models. However, there are exceptions where the effective application of certain metrics hinges on the availability of specific types of data. For instance, coverage-based metrics require knowledge of item categories, whereas distance-based metrics rely on the availability of item embeddings, illustrating that while metrics are broadly applicable, their utility may vary depending on certain specifics.}

\new{When it comes to diversity approaches, our discussion is primarily centered on offline approaches due to their prevalent use in current research and applications. These approaches can be broadly categorized into pre-processing, in-processing, and post-processing strategies, each with its unique considerations and potential exceptions regarding their integration with different recommendation models. Pre-processing strategies, such as defining user types, establishing sampling strategies, and setting ground truth labels, offer broad applicability but vary in necessity depending on the model's loss function. For instance, pair-wise loss models~\cite{ZhengGCJL21DGCN-pre} may benefit more from sampling strategies than point-wise loss models, yet all models can leverage user types and ground truth labels to enhance recommendation precision. In-processing approaches, characterized by their ability to treat diversity either as regularization or directly within the scoring function, show flexibility across models. 
Although intuitively aligned with list-wise loss models due to their holistic assessment of recommendation lists, these approaches can also be adapted for pair-wise and point-wise models, demonstrating the potential for wide applicability through creative adaptations like constructing lists from pair-wise or point-wise scores to apply diversity constraints~\cite{WasilewskiH16Diverse_LTR, LiZZZL17DiverseMF}. 
Post-processing strategies stand out for their universal compatibility, enabling the integration of diversity enhancements after model training, thereby maintaining their efficacy across all types of models.}

\new{The emergence of Large Language Models (LLMs) introduces new considerations for the application of diversity approaches. LLMs, with their advanced natural language processing capabilities, present unique challenges and opportunities for integrating diversity metrics and approaches. Diversity metrics are still applicable, while the direct application of diversity approaches to LLM-based recommendation systems may not be that straightforward.
For instance, applying in-processing approaches to models utilizing LLMs for recommendations via prompt engineering is challenging, due to their unique architectures and data representation~\cite{chen2023palr, radlinski2022natural}.
However, when LLMs are used to augment existing recommendation frameworks by enhancing data quality or providing auxiliary information~\cite{wei2024llmrec,ren2024representation}, the potential for applying diversity approaches remains viable.}

\new{In summary, while diversity metrics and approaches generally maintain a high degree of model-agnosticism, enabling their application across a spectrum of models, certain exceptions and considerations must be acknowledged. This includes the unique challenges posed by the integration of LLMs into the recommendation landscape, underscoring the need for adaptive and flexible strategies to ensure the effective incorporation of diversity considerations in modern search and recommender systems.}
\vspace{-2mm}
\section{Openness and Future Directions}
\label{sec:future}

Researchers have realized the importance of improving diversity in retrieval systems and have started the exploration. 
However, we argue that there still exists openness in this area. 
In this section, we discuss a number of open challenges and point out some future opportunities in an effort to encourage additional research in this field.

\vspace{-3mm}
\subsection{Time Dependency}
Existing research on diversity-aware retrieval systems focuses primarily on a single time point without taking a continuous time span into account.
In real-world systems, however, time plays an important role in the study of user behaviors and intentions, as humans may require varying degrees of diversity at different stages of their interaction with the system.
We argue that an intriguing future research direction is to investigate how to ensure personalized and time-dependent diversity in a continuous learning setting in which data arrive in a time-series fashion.
For instance, when a new user first joins a system, it is reasonable for the algorithm to display more diverse results in order to help the user better explore her interests.
As more data is collected about the user's interaction with the system, the algorithm should be able to adjust itself to adaptively balance relevance and diversity in order to not only provide items that the user likes based on the user's past preferences, but also show serendipity to the user at some point in order to attract and retain the user.

\vspace{-2mm}
\subsection{Direct Optimization of Metrics}
One of the challenges in enhancing the diversity of search results in retrieval systems is that some metrics are difficult to optimize directly.
Although methods have been proposed to make some metrics differentiable (e.g., $\alpha$-nDCG, Gini Index)~\cite{WuCZZ09SmoothDCG, DoU22Gini}, most metrics, such as coverage-based metrics and SD Index, are difficult to optimize directly.
This hinders the capacity of in-processing methods to achieve trade-offs between diversity and other metrics.
Exploring a more general method for differentiating these metrics for end-to-end training could be an intriguing line of research.

\vspace{-2mm}
\subsection{Diversity in Explainability}
While much of the diversity-aware research on search and recommendation concentrates on presenting a diverse item list to users, diversity can also pertain to other dimensions like explainability, equally important for user retention and satisfaction. 
For example, it is not ideal if a recommender system always explains to a user as \textit{``based on your previous history''} or \textit{``similar users also like...''}. 
This research direction has received scant attention, with only a few works exploring this area~\cite{li2020directional, DBLP:conf/sigir/BalloccuBFM22}. 
We think it is intriguing to explore what user and item features cause varying degrees of diversity in output lists, as this understanding can guide the generation of diverse user explanations.

\vspace{-2mm}
\subsection{Multi-Stakeholder Trade-offs}
Managing multi-stakeholder interests in search and recommendation systems presents a complex challenge, involving balancing the needs and preferences of users, content creators, platform owners, \new{society, etc~\cite{abdollahpouri2020multistakeholder}.} Currently, there is a dearth of exploration in how to simultaneously ensure relevance, diversity, and fairness among these entities. For example, while a user might desire a highly personalized and diverse set of recommendations, content creators could desire a broad distribution of their content to reach a wider audience. Similarly, platform owners might want to maximize user engagement and the time spent on their platform while maintaining a fair playing field for all content creators. 
\new{
Current literature on diversity in search and recommendation, however, predominantly centers on users or content creators, rather than embracing a system-wide perspective. 
Consequently, this has resulted in a scarcity of comprehensive reviews on diversity that incorporate these broader stakeholder perspectives. 
We observe that there exist strong relations among different stakeholders, for instance, promoting diversity at the individual level (to meet user satisfaction) and at the system level (to cater to content creators) can be seen as enhancing the overall diversity of the system, which in turn satisfies platform owners. 
Additionally, when including ``society'' as a stakeholder, the social welfare metrics discussed in Sec.~\ref{sec:social_welfare_metric}, such as the Simpson’s Diversity Index and the Gini Index, become crucial targets for optimization.}
Future research in this area could investigate mechanisms to navigate these trade-offs and model the dynamics between multiple stakeholders, providing a more holistic approach to designing diversity-aware retrieval systems.

\vspace{-2mm}
\subsection{Diversity in Multi-Modal Recommendation}
Multi-modal recommendation systems, such as those integrating text, images, and audio, pose unique challenges and opportunities for diversity. So far, most diversity-aware systems have focused on single-mode recommendation, such as text-only or image-only recommendations. However, as our digital world becomes increasingly multi-modal, the concept of diversity must expand to consider how different modalities contribute to or detract from diversity. This can be a complex task, given the distinct characteristics and inherent diversity within and across different modalities. For instance, diversity in text might relate to content topic or writing style, while diversity in images might be related to color, style, or content theme. In audio recommendations, diversity could be associated with various factors like genre, artist, or mood. Identifying how to effectively measure and optimize diversity in a multi-modal context is an open challenge that requires further investigation. Future research in this area can bring new insights into how different modalities interact and how they can be combined to create more diverse and enriching user experiences.

\vspace{-2mm}
\subsection{\new{Relation to Other Metrics}}
\new{Diversity cannot be viewed in isolation; it is interconnected with other important metrics. In our survey, we touch upon how diversity relates to relevance and novelty. Another critical metric that is often discussed in conjunction with diversity is fairness~\cite{gao2021addressing, verma2020facets}. Enhancing diversity can align with efforts to increase fairness in recommendations, contributing to a more balanced and equitable distribution of content across users, which may further benefit the entire society. Exploring the trade-offs between diversity and fairness metrics presents a compelling future direction.}

\new{Moreover, with the rise of LLMs, the evaluation of diversity in natural language generation has garnered more interest. For example, \citet{tevet2020evaluating} provide a summary of diversity metrics in this domain and assess their effectiveness. There is potential for overlap between diversity measures in natural language generation and those in search and recommendation systems, particularly as the use of LLMs in search and information retrieval becomes more prevalent. Adopting these metrics in search and recommendation contexts could offer a richer evaluation framework, allowing for insights from diverse angles.}





\section{Conclusion}
In this survey, we introduce the foundations, definitions, metrics, and approaches of diversity in retrieval systems from the perspective of search and recommendation.
We begin the survey with a introduction of why diversity is important in retrieval systems for the benefit of multiple stakeholders.
To help better understand the diversity concepts, we summarize the different diversity concerns in search and recommendation, highlighting their connection and distinctions. 
For the main body of the survey, we provide a unified taxonomy to classify the metrics and approaches of diversification in both search and recommendation.
To close the survey, we discuss the open research questions of diversity-aware research in retrieval systems in the hopes of inspiring future innovations and encouraging the deployment of diversity in real-world systems.

\bibliographystyle{IEEEtranN}
\bibliography{reference}

\begin{IEEEbiography}[{\includegraphics[width=1.0\columnwidth]{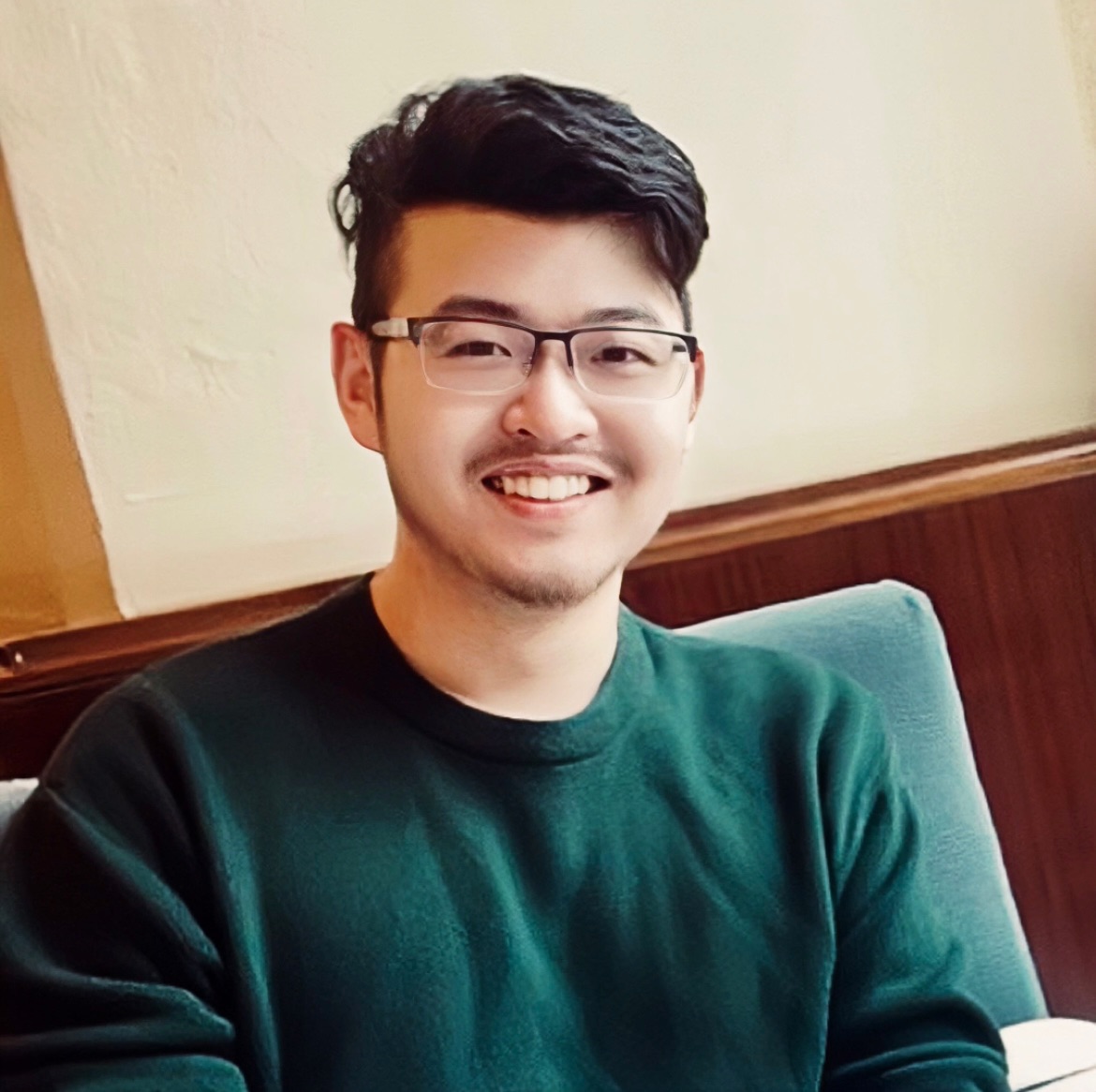}}]{Haolun Wu} is a Ph.D. candidate in Computer Science at Mila - Quebec AI Intitute and McGill University.
His research interests include in information retrieval, recommender systems, knowledge modeling, LLMs, etc.
He has published papers on ICLR, SIGIR, AAAI, CIKM, TOIS, ICDE, CHI, etc.
He is a student researcher at Google Research and Microsoft Research.
He holds the Borealis AI Fellowship 2023-24.
\end{IEEEbiography}

\vspace{-15mm}
\begin{IEEEbiography}[{\includegraphics[width=1.0\columnwidth]{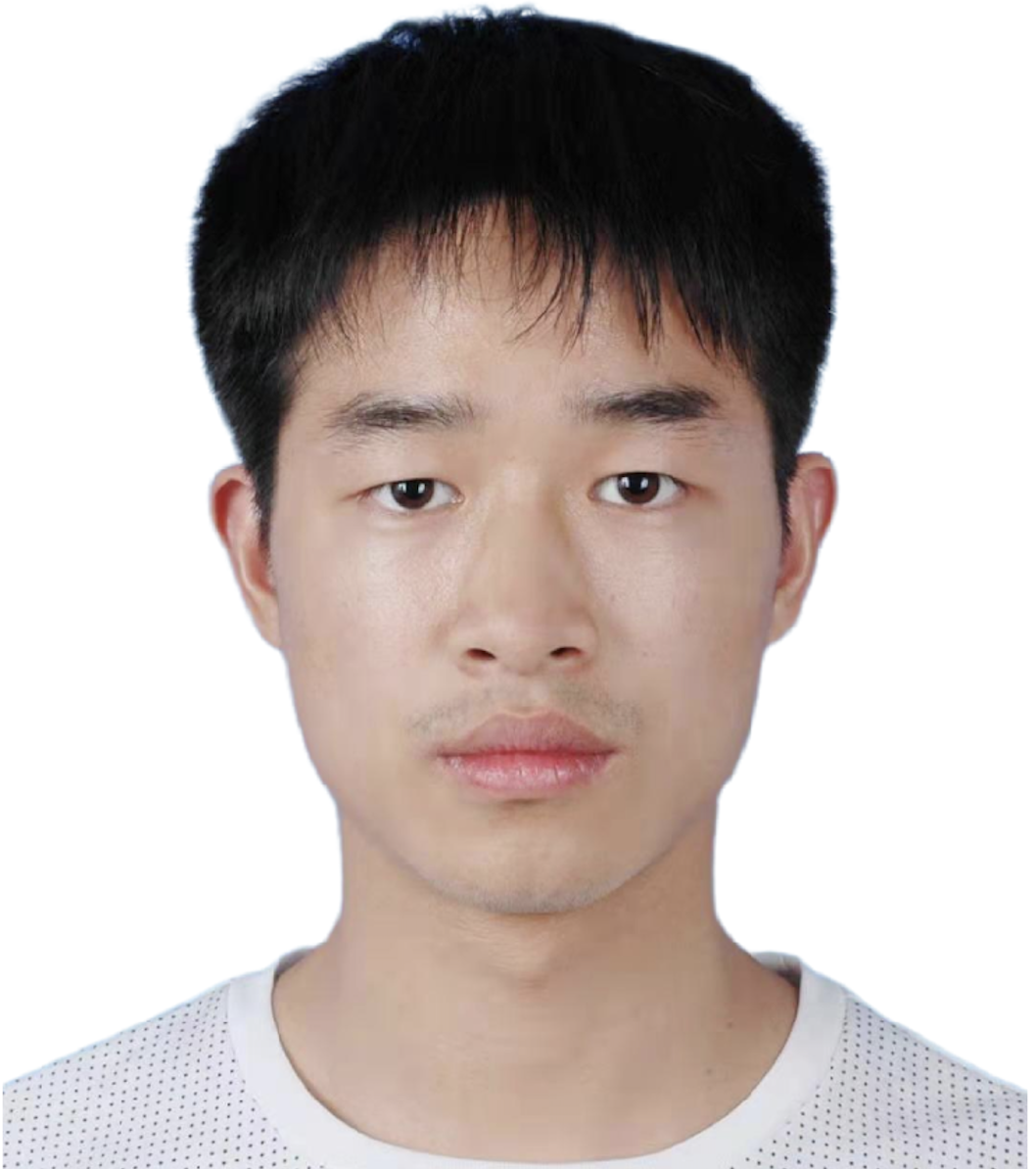}}]{Yansen Zhang} is a Ph.D. candidate in Computer Science at City University of Hong Kong. 
He received his B.S. and M.S. degrees in Software Engineering from NEU and SYSU, China, in 2019 and 2022, respectively.
His research interests include diversified recommendation and explainable recommendation.
He has published papers on SIGIR and ICONIP.
\end{IEEEbiography}

\vspace{-15mm}
\begin{IEEEbiography}[{\includegraphics[width=1.0\columnwidth]{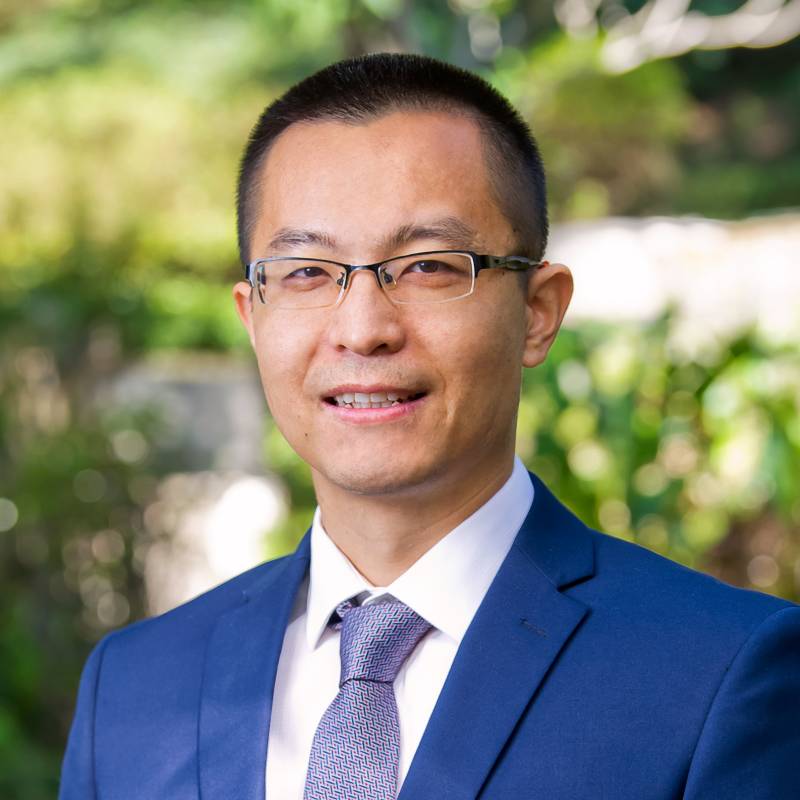}}]{Chen Ma} is currently an Assistant Professor in the Department of Computer Science, City University of Hong Kong. His research interests lie in the theory of data mining and machine learning and their applications in recommender systems, knowledge graphs, social computing, and social good. He received his B.S. and M.S. degrees in Software Engineering from Beijing Institute of Technology in 2013 and 2015, respectively. He received his PhD degree in Computer Science from McGill University.
\end{IEEEbiography}

\vspace{-15mm}
\begin{IEEEbiography}
[{\includegraphics[width=1.0\columnwidth]{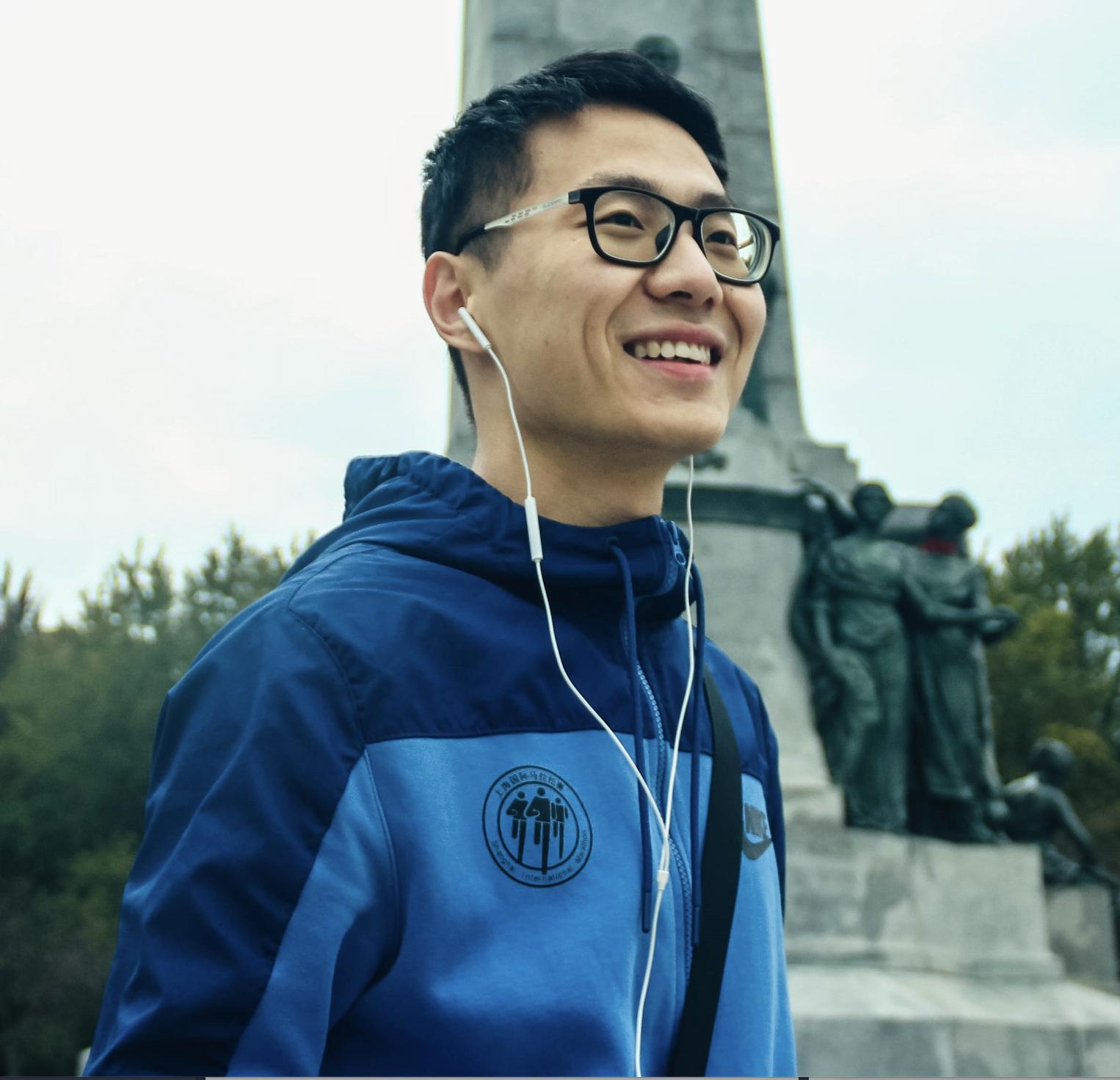}}]{Fuyuan Lyu} is a Ph.D. candidate in Computer Science at McGill University. He obtained his B.S. degree from Shanghai Jiao Tong University with Zhiyuan honor degree. His research interest lies in the intersection between AutoML and IR. He has published papers on NeurIPS, WWW, KDD, ICDE, CIKM, etc. He interned and collaborated with Huawei Noah's Ark Lab and Tencent Financial Technology. He previously worked as a research assistant at Nanyang Technological University and Shanghai Jiao Tong University.
\end{IEEEbiography}

\vspace{-15mm}
\begin{IEEEbiography}
[{\includegraphics[width=1.0\columnwidth]{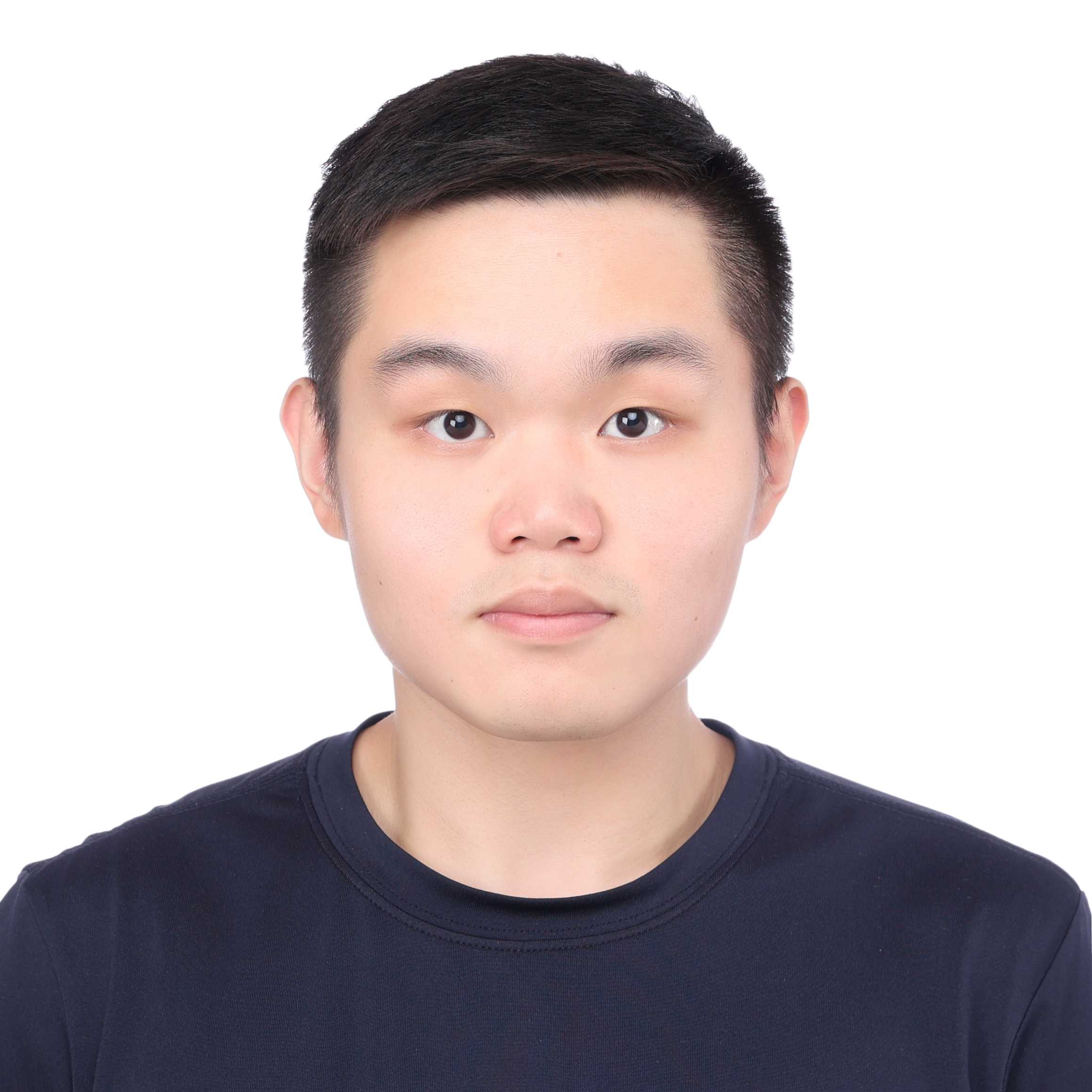}}]{Bowei He} is a Ph.D. candidate in Computer Science at City University of Hong Kong. His research focuses on advanced ML (e.g., RL, continual learning, and Auto-ML) and causal inference for data mining and information retrieval. He has published papers on WWW, NeurIPS, ICDE, CVPR, ICCV, CIKM, ICDM, UAI, SDM, ICME, ICASSP, IROS, etc. He interned and collaborated with Didi AI Labs, Huawei Noah's Ark Lab, and Tencent Financial Technology.
\end{IEEEbiography}

\vspace{-15mm}
\begin{IEEEbiography}
[{\includegraphics[width=1.0\columnwidth]{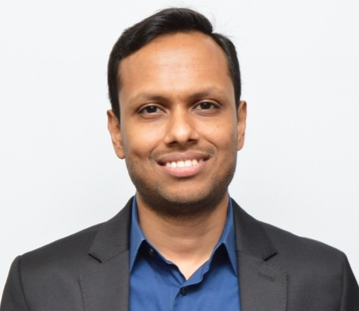}}]{Bhaskar Mitra} is a Principal Researcher in the Collaborative Intelligence group at Microsoft Research. His research interests lie at the intersections of deep learning, information retrieval, knowledge bases, benchmarking and evaluation, and FATE (Fairness, Accountability, Transparency, and Ethics). He co-created the MS MARCO benchmark and co-organizes the TREC Deep Learning and Tip-of-the-Tongue Tracks. He received his Ph.D. in Computer Science from University College London.
\end{IEEEbiography}

\vspace{-15mm}
\begin{IEEEbiography}[{\includegraphics[width=1.0\columnwidth]{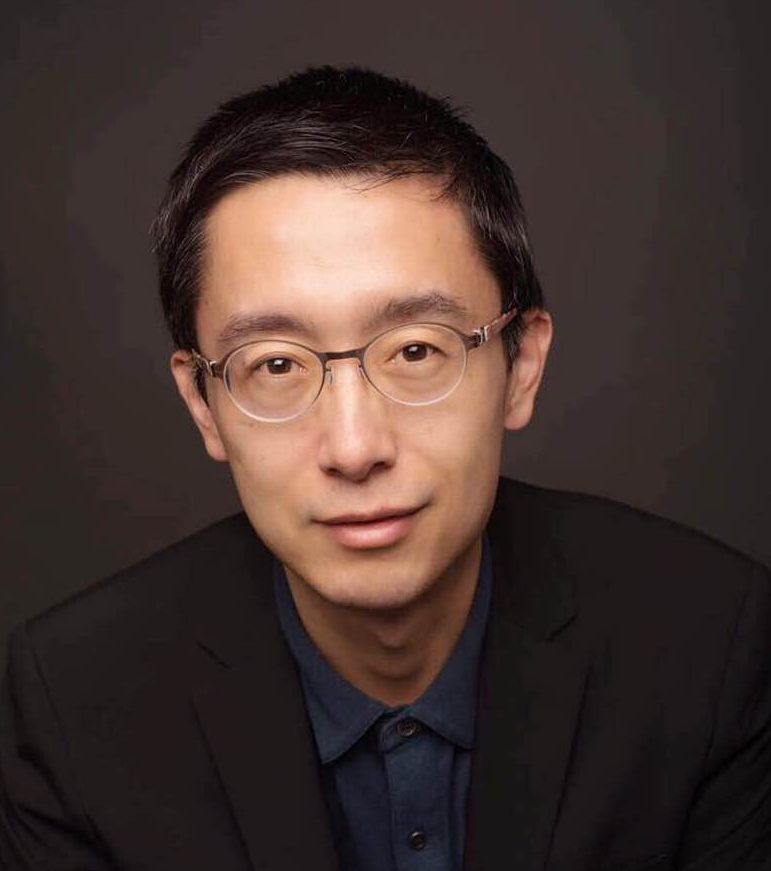}}]{Xue Liu (Fellow, IEEE)} is a Professor and a William Dawson Scholar in the School of Computer Science, McGill University.
He is a Fellow of the Canadian Academy of Engineering (FCAE) and a Fellow of IEEE (FIEEE). He is an associate member of Mila, and also the Chief Scientist and Co-Director of Samsung AI Center Montreal.
He was also the Samuel R. Thompson Associate Professor with the University of Nebraska-Lincoln and HP Labs, Palo Alto, USA. 
He received the B.S. degree in mathematics and the M.S. degree in automatic control from Tsinghua University in 1996 and 1999, respectively, and the Ph.D. degree in computer science from the University of Illinois at Urbana–Champaign in 2006. 
He has over 150 published research papers. His areas of interest encompass computer networks and communications, smart grid, real-time and embedded systems, cyber-physical systems, data centers, and applied ML.
\end{IEEEbiography}

\end{document}